\documentclass[useAMS,usenatbib]{iopart}
\usepackage{graphicx}
\usepackage{amssymb}
\usepackage{color}

\begin{document}

\title[SPHINCS\_BSSN: SPH in dynamical spacetimes]{SPHINCS\_BSSN: A general relativistic Smooth Particle Hydrodynamics code for dynamical spacetimes}
\author[Rosswog \&Diener]{
 S. Rosswog$^1$  \& 
   P. Diener$^{2,3}$}
$^1$ Astronomy and Oskar Klein Centre, Stockholm University, AlbaNova, SE-10691 Stockholm, Sweden\\
$^2$ Center for Computation \& Technology, Louisiana State University, Baton Rouge, LA 70803, USA\\
$^3$ Department of Physics \& Astronomy, Louisiana State University, Baton Rouge, LA 70803, USA

\newcommand{\nifs}{\ensuremath{^{56}\mathrm{Ni}}}
\newcommand{\Nifs}{\ensuremath{^{56}\mathrm{Ni}} $\;$}
\def\paren#1{\left( #1 \right)}
\def\Mesz{M\'esz\'aros~}
\def\Pacz{Paczy\'nski~}
\def\Kluz{Klu\'zniak~}
\def\p{\partial}
\def\msun{M$_{\odot}$}
\def\Msun{M$_{\odot}$ }
\def\be{\begin{equation}}
\def\ee{\end{equation}}
\def\bi{\begin{itemize}}
\def\i{\item}
\def\ei{\end{itemize}}
\def\ben{\begin{enumerate}}
\def\een{\end{enumerate}}
\def\bea{\begin{eqnarray}}
\def\eea{\end{eqnarray}}
\def\bt{\begin{tabbing}}
\def\et{\end{tabbing}}
\def\gcc{gcm$^{-3}$}
\def\Gcc{gcm$^{-3} \;$}
\def\ccm{cm$^3$}
\def\edo{\end{document}}
\def\red{\textcolor{red}}
\def\blue{\textcolor{blue}}
\def\green{\textcolor{green}}
\def\ye{$Y_e$}
\def\Ye{$Y_e$ }
\def\rs{$R_{\rm S}$}
\def\Rs{$R_{\rm S}$ }
\def\noin{\noindent}
\def\QIone{\mathcal{Q}_{\; \; 1}}
\def\QItwo{\mathcal{Q}_{\; \; 2}}
\def\QIthree{\mathcal{Q}_{\; \; 3}}
\def\QIfour{\mathcal{Q}_{\; \; 4}}
\newcommand{\ud}{\mathrm{d}}
\newcommand{\sn}{\ensuremath{\mathrm{sn}}}
\newcommand{\cn}{\ensuremath{\mathrm{cn}}}
\newcommand{\dn}{\ensuremath{\mathrm{dn}}}
\newcommand{\sech}{\ensuremath{\mathrm{sech}}}
\newcommand{\nd}{\noindent}
\newcommand{\enth}{\mathcal{E}}
\newcommand{\carb}{$^{12}$C}
\newcommand{\Carb}{$^{12}$C $\;$}
\newcommand{\ox}{$^{16}$O}
\newcommand{\Ox}{$^{16}$O $\;$}

\def\peak{\rm peak}
\def\ji{\varphi}
\def\rin{\mathcal{I}}
\def\x{{\bf x}}
\def\s{{\bf s}}
\def\k{{\bf k}}
\def\n{{\bf n}}
\def\f{{\bf f}}
\def\pd{\partial}
\def\fa{f_{\alpha}(\x,\p,t)}
\def\kabs{\kappa^{\rm abs}_\nu}
\def\ksca{\kappa^{\rm sca}_\nu}
\def\Flux{{\bf F}_\nu}
\def\Frad{{\bf F}_{\rm rad}}
\def\Pressure{{\mathsf P}_\nu}
\def\kB{k_B}
\def\tauph{\tau_{\it ph}}
\def\taumu{\tau_\mu}
\def\MSun{M_{\astrosun}}

\newcommand{\apj}{Ap. J., }
\newcommand{\apjl}{Ap. J. Lett., }
\newcommand{\apjs}{Ap. J. Supp., }
\newcommand{\physrep}{Phys. Rep., }
\newcommand{\mnras}{Mon. not. RAS., }
\newcommand{\apss}{Astroph. \& Space Sci., }
\newcommand{\aap}{Astron. \& Astrophys., }
\newcommand{\prc}{Phys. Rev. C.,} 
\newcommand{\prd}{Phys. Rev. D.,} 
\newcommand{\nat}{Nature,}
\newcommand{\pasp}{Publications of the Astronomical Society of the Pacific,}

\def\M4{M$_4$}
\def\M6{M$_6$}
\def\lcm6{LCM$_6$}
\def\qcm6{QCM$_6$}
\def\whn{W$_{\rm H,n}$}
\def\wh3{W$_{\rm H,3}$}
\def\wh4{W$_{\rm H,4}$}
\def\wh5{W$_{\rm H,5}$}
\def\wh6{W$_{\rm H,6}$}
\def\wh7{W$_{\rm H,7}$}

\def\divv{\nabla \cdot \vec{v}}
\def\vv{\vec{v}}
\def\dx{\Delta x}

\newcommand{\ve}[1]{\boldsymbol{#1}}

\def\ed{\dot{\epsilon}}
\def\vl{v_\lambda}
\def\vort{\vec{\mathcal{W}}}

\def\ent{\mathcal{E}} 

\newcommand{\SpB}{\texttt{SPHINCS\_BSSN}\enskip}
\newcommand{\spB}{\texttt{SPHINCS\_BSSN}}
\newcommand{\Ma}{\texttt{MAGMA2}\enskip}
\newcommand{\ma}{\texttt{MAGMA2}}
\newcommand{\ahf}{\texttt{AHFinderDirect}}
\newcommand{\Ahf}{\texttt{AHFinderDirect}\enskip}
\newcommand{\mcl}{\texttt{McLachlan}}
\newcommand{\Mcl}{\texttt{McLachlan}\enskip}
\newcommand{\tlg}{\tilde{\gamma}}
\newcommand{\emfp}{e^{-4\phi}}
\newcommand{\emfpW}{W^2}
\newcommand{\tlG}{\tilde{\Gamma}}
\newcommand{\tlGn}{\tilde{\Gamma}_{\mathrm{(n)}}}
\newcommand{\tlGmixed}[2]{\tlG_{#1}^{\;\;\;#2}}
\newcommand{\tlA}{\tilde{A}}
\newcommand{\tlR}{\tilde{R}}
\newcommand{\xt}{\tilde{x}}
\newcommand{\dt}[1]{\partial_t #1}
\newcommand{\pdu}[3]{\partial_{#3} #1^{#2}}
\newcommand{\pdl}[3]{\partial_{#3} #1_{#2}}
\newcommand{\pdpdu}[4]{\partial_{#3}\partial_{#4} #1^{#2}}
\newcommand{\pdpdl}[4]{\partial_{#3}\partial_{#4} #1_{#2}}
\newcommand{\upwindu}[3]{\beta^{#3}\bar{\partial}_{#3} #1^{#2}}
\newcommand{\upwindl}[3]{\beta^{#3}\bar{\partial}_{#3} #1_{#2}}

\begin{abstract}
We present a new methodology for simulating self-gravitating general-relativistic 
fluids. In our approach the fluid is modelled by means of Lagrangian particles in the framework of
a general-relativistic (GR) Smooth Particle Hydrodynamics (SPH) formulation,
while the spacetime is evolved on a mesh according to
the BSSN formulation that is also frequently used in Eulerian GR-hydrodynamics. To the best of our 
knowledge this is the first Lagrangian fully general relativistic hydrodynamics code (all previous 
SPH approaches used approximations to GR-gravity).
A core ingredient of our particle-mesh approach is the coupling between the gas 
(represented by particles) and the spacetime (represented by a mesh) for which we have 
developed a set of  sophisticated interpolation tools that are inspired by other particle-mesh approaches,
in particular by vortex-particle methods. One advantage of splitting the methodology
between matter and spacetime is that it gives us more freedom  in choosing the resolution, 
so that --if the spacetime is smooth enough-- we obtain good results already with a moderate number 
of grid cells and can focus the computational effort on the simulation of the matter. Further advantages
of our approach are the ease with which ejecta can be tracked and the fact that the neutron star surface
remains well-behaved and does not need any particular treatment.
In the hydrodynamics part of the code we  use a number of techniques that are 
 new to SPH, such as reconstruction, slope limiting and steering dissipation by monitoring entropy conservation. 
 We describe here in detail the employed numerical methods and demonstrate the code performance in a 
 number of benchmark problems ranging from shock tube tests, over Cowling approximations 
to the fully dynamical evolution of neutron stars in self-consistently evolved spacetimes. 
\end{abstract}


\section{Introduction}
The first detection of gravitational waves (GWs) from a merging binary black hole \cite{abbott16a} 
opened up the sky for a side of the Universe that was previously invisible.
Through this milestone event, gravitational wave detections have become an active part of observational 
astronomy. The next watershed event followed soon after: in August 2017  a 
binary neutron star merger was detected \cite{abbott17b,abbott17c}, first via gravitational and then via
electromagnetic (EM) waves.
The gravitational waves provided stringent limits on the tidal deformability of the neutron stars and thus
constrained the properties of matter at supra-nuclear densities \cite{abbott17b}. The detection 
of a short GRB 1.7 s after the  GW-peak \cite{abbott17d,goldstein17,savchenko17,troja17,hallinan17,kasliwal17,mooley18} 
confirmed the long-standing expectation \cite{eichler89} that neutron star mergers are indeed
GRB progenitors and the time delay between both signals provided the tightest constraints so far on 
GWs propagating, with an enormous precision, at the speed of light \cite{abbott17d}. The merger event also allowed for an 
independent determination of the Hubble parameter \cite{abbott17a}.
The early UV, optical and IR radiation that were detected within about one day after the GW-peak,
were consistent with the expectations for transients that are powered by the radioactivity from freshly
synthesized ``rapid neutron capture'' or ``r-process'' material, so-called ``macronovae'' \cite{kulkarni05} or ``kilonovae'' \cite{metzger10b}. 
In particular the bolometric luminosity was consistent with being powered by a broad distribution  of r-process elements 
\cite{rosswog18a,metzger19a}, thereby confirming neutron star mergers as a major
cosmic r-process production site \cite{lattimer74,lattimer77,eichler89,rosswog99,freiburghaus99b},
see \cite{cowan20} for a recent, extensive review.
The spectral evolution from the blue ($\sim$ day)  to red emission ($\sim$ week) 
suggests that matter with a broad range of electron fractions was ejected, extending from the very low values
in the original neutron star, $Y_e\sim0.05$, to values exceeding $Y_e^{\rm crit}\approx 0.25 $. 
At this critical value the matter composition changes abruptly \cite{korobkin12a,lippuner15,kasliwal19} 
and for larger values the ejecta contain no more lanthanides and actinides which are major
opacity sources \cite{kasen13a,tanaka13a,tanaka20a}.
Since the original neutron star only contains tiny amounts ($\sim 10^{-5}$ \msun) of matter with
$Y_e>Y_e^{\rm crit}$, this demonstrates that we have been witnesses to weak interactions at work. \\
While all of the above were major strides forward for individual topics and questions, this observation 
was also a spectacular reminder of the multi-physics nature of neutron star mergers. Major breakthroughs
were possible since the signatures of bulk flows in curved spacetime, gravitational waves, were
detected in concert with the signature of relatively small amounts of mass ($\sim 10^{-2}$ \msun)
whose nuclear (composition, heating rates) and atomic properties (line opacities) shape 
the electromagnetic emission. The event also emphasized  that, for reliable multi-messenger 
modelling, all the fundamental forces of nature need to be included together with a broad range 
of length (from $\sim 10^4$ cm for the pressure scale height within a neutron star to $\sim 10^{15}$ cm 
for the ejecta size at the emission peak) and time scales (from sub-millisecond dynamical time scales of 
neutron stars to $\sim 1$ week for the EM emission). Apart from emphasizing the need for a broad range
of physics ingredients, this event also illustrates how demanding the numerical modelling of
such mergers is.
\\
As outlined above, both the high density bulk-flows (for GWs) and the small amounts of 
low-density ejecta (for the EM signal) need to be faithfully modelled.
To date all fully relativistic hydrodynamics approaches are based on Eulerian hydrodynamic 
formulations, see e.g. \cite{alcubierre08,baumgarte10,rezzolla13a,shibata16}. While these
methods have delivered a plethora of important results \cite{baiotti17,duez19a,shibata19a}, they
are also facing some challenges. For example, a neutron star surface is a region that is notoriously 
difficult to handle. Most Eulerian Numerical Relativity codes cannot handle regions with true vacuum
in simulations that also involve matter\footnote{But see
\cite{poudel20} for recent progress.} and therefore the neutron
stars are embedded in a non-zero density ``atmosphere'' which can lead to failures 
in recovering the primitive variables and to an effective reduction of the convergence order \cite{schoepe18}.
Moreover, the  small amounts of ejecta have to escape against the (hopefully negligible)
resistance of this background medium. Another challenging issue for Eulerian hydrodynamics is that
advection is not exact and following ejecta to large distances, where the hydrodynamic resolution 
usually deteriorates, can become difficult. \\
Lagrangian methods offer an interesting alternative, since they can make advection exact and vacuum
corresponds to true absence of matter, but to date no fully relativistic Lagrangian hydrodynamics 
code is available. Commonly used Lagrangian methods include Smooth Particle Hydrodynamics (SPH)
\cite{monaghan05,rosswog09b,springel10a,price12a,rosswog15c}, finite volume approaches formulated
on moving meshes based on Voronoi tesselations \cite{springel10b,duffell11} or finite volume methods
that are based on overlapping, spherical particles \cite{gaburov11,hopkins15a,hubber18}. Such methods have the
advantage that they are not restricted by a predescribed  mesh geometry and they are very accurate in 
terms of advection.\\
SPH methods based on Newtonian gravity (plus GW back reaction forces) have been used early on 
to model compact mergers with nuclear matter and neutrino effects 
\cite{rosswog99,rosswog02a,rosswog03a,rosswog03c}. There are also post-Newtonian SPH formulations
\cite{faber00,ayal01,faber01} that are based on the work of \cite{blanchet90}, but the practical applicability of these 
approaches to neutron stars has remained very limited. The closest approximation to general relativistic
strong field gravity to date in SPH has been the conformal 
flatness approximation \cite{oechslin02,oechslin04,faber06a,faber06b,bauswein10a}, 
but to date no Lagrangian hydrodynamics code exists that self-consistently evolves matter and spacetime.\\
In this paper, we describe the first such approach, which has been implemented in the new code
 \SpB (``Smoothed Particle Hydrodynamics In Curved Spacetime using BSSN''). We solve the relativistic hydrodynamics equations
by means of freely moving SPH particles and, based on their energy-momentum tensor, evolve the 
spacetime according to the BSSN formulation  \cite{nakamura87,shibata95,baumgarte99}. 
Our paper is structured as follows. In Sec.~\ref{sec:meth} we describe first how we evolve the relativistic 
fluid,  then how we treat the spacetime and, finally, how we couple both together. Sec.~\ref{sec:tests} is dedicated to  a 
number of benchmark tests and we  conclude with a summary in Sec.~\ref{sec:summary}.

\section{Methodology}
\label{sec:meth}

\subsection{Broad-brush overview over our algorithm}
Since a number of rather technical steps are involved, we will first give
a broad-brush overview over our algorithm before we explain the details
of the involved ingredients.
We use a hybrid approach where we follow the hydrodynamic evolution of matter by means of Lagrangian particles,
as described in Sec.~\ref{sec:hydro}, while the spacetime is evolved via the BSSN approach \cite{shibata95,baumgarte99} 
using a Cartesian mesh, see Sec.~\ref{sec:spacetime_evolution}. The particles and the mesh need to communicate:
\bi
\item {\em particles need from mesh}: the metric $g_{\mu\nu}$ and the "metric acceleration terms" for the momentum and energy 
equations, Eqs.~(\ref{eq:dSdt_metric}) and (\ref{eq:dedt_metric}), at the particle locations,
\item {\em mesh needs from particles}: the energy-momentum tensor $T_{\mu\nu}$ for the source terms in BSSN, 
see Eqs.(\ref{eq:bssn_ev_start} - \ref{eq:BSSN_Si}), at the grid points.
\ei 
This communication between the particles and the mesh is a crucial ingredient of our approach, it is described 
in detail in Sec.~\ref{sec:particle_mesh}.\\
Assume that we have a consistent set of initial conditions both for the
spacetime and the matter ("hydrodynamic") variables. For the sake of a
compact notation, we will collectively refer to the hydrodynamic evolution variables 
as $\vec{Y}^{\rm hyd}$, while the spacetime evolution variables are denoted as
$\vec{Y}^{\rm met}$,  together they form the vector 
$\vec{Y}= (\vec{Y}^{\rm hyd},\vec{Y}^{\rm met}$) that is integrated forward in time. As will be
explained in more detail below, our hydrodynamic variables consist of a
baryon number density $N^\ast$, a momentum variable $S_i$ and an
energy variable $e$, see Sec.~\ref{sec:hydro}, while our spacetime variables
are the standard BSSN variables, see Sec.~\ref{sec:spacetime_evolution}.
The vector $\vec{Y}$ is integrated forward in time from $t^n$ via an optimal 3rd order
TVD Runge-Kutta approach \cite{gottlieb98} to a time $t^{n+1}$:
\bea
\vec{Y}^{(1)} &=& \vec{Y}^{n} + \Delta t \; L(\vec{Y}^{n})\\
\vec{Y}^{(2)} &=& \frac{1}{4}[ 3 \vec{Y}^{n}  + \vec{Y}^{(1)} +  \Delta t \; L(\vec{Y}^{(1)}) ]\\
\vec{Y}^{n+1} &=& \frac{1}{3} [ \vec{Y}^{n} + 2  \vec{Y}^{(2)} + 2 \Delta t \; L(\vec{Y}^{(2)}) ],
\eea
where $L(\vec{Y})$ denote the derivatives evaluated at $\vec{Y}$.\\
The workflow within one Runge-Kutta sub-step is the following:
\ben
\i convert the BSSN variables to the physical metric, see Eq.~(\ref{eq:phys_metric})
\i map physical metric, $g_{\mu\nu}$, from the mesh to the particle positions, see Mesh-to-Particle step in Sec.~\ref{sec:particle_mesh}
\i update the tree-structure for neighbour search and update each particle's smoothing length, see Sec.~\ref{sec:SPH_kernel}
\i calculate the density variable $N^\ast$, see Eq.~(\ref{eq:N_sum})
\i recover the physical variables (specific energy per baryon $u$, local rest frame baryon number density $n$, and velocities $v^i$) from the numerical variables $N^\ast$, $S_i$ and $e$, see Sec.~\ref{sec:recovery}
\i map the energy-momentum tensor $T_{\mu \nu}$ from the particle positions to the mesh, see the Particle-to-Mesh step in Sec.~\ref{sec:particle_mesh}
\i calculate the time derivatives of the BSSN variables, see Eq.~(\ref{eq:bssn_ev_start}) to Eq.~(\ref{eq:bssn_ev_end})
\i calculate the time and spatial derivatives of the physical matric from the
BSSN variables by applying the chain rule to Eq.~(\ref{eq:phys_metric})
\i map $\frac{\sqrt{-g}}{2}\frac{\partial g_{\mu\nu}}{\partial x^{\mu}}$ from the mesh to the particle positions, see Mesh-to-Particle step in Sec.~\ref{sec:particle_mesh}
\i calculate the metric acceleration terms, Eqs.~(\ref{eq:dSdt_metric}) and (\ref{eq:dedt_metric}), on the particles
\i calculate the time derivatives of the hydrodynamic variables, $dS_i/dt$ and $de/dt$, see Eqs.(\ref{eq:dSdt_full}) and (\ref{eq:energy_equation})
\i update the time step as the minimum of the hydrodynamic and the BSSN time step, $\Delta t= {\rm min}(\Delta t_{\rm hyd},0.35\, \Delta/c)$, where $\Delta t_{\rm hyd}= 0.2\, {\rm min}_a (h_a/c)$ and $h_a$ is the smoothing length of  particle $a$ and $\Delta$ is the grid spacing.
\een
After this short overview over the workflow, we will describe the employed ingredients in more detail in the following.

\subsection{Hydrodynamics}
\label{sec:hydro}
\subsubsection{Non-dissipative SPH}
\label{sec:ideal_hydro}
SPH in Newtonian, special- and general relativistic form can be  elegantly derived from a discretized fluid Lagrangian  
\cite{monaghan01,rosswog09b,rosswog10a,rosswog10b,rosswog15b}. 
We use  $c=1$ and metric signature ($-,+,+,+$), greek indices run from 0..3 and latin indices from 1..3.
 Contravariant spatial indices of a vector
quantity $w$ at particle $a$ are denoted as $w^i_a$, while covariant ones will be written as
$(w_i)_a$.\\
Here we only briefly sketch the derivation of the equations that we are using, the detailed steps can be found
in Sec. 4.2 of \cite{rosswog09b}\footnote{The extension of this derivation to the case including 
(small) terms from derivatives of the SPH kernels with respect to the smoothing lengths can be found
in \cite{rosswog10a}.}.
The line element and proper time are given by $ds^2= g_{\mu \nu} \, dx^\mu \, dx^\nu$ and
$d\tau^2= - ds^2$ and the line element in a 3+1-split of spacetime reads
\be
ds^2= -\alpha^2 dt^2 + \gamma_{ij} (dx^i + \beta^i dt) (dx^j + \beta^j dt),
\ee
where $\alpha$ is the lapse function, $\beta^i$ the shift vector and $\gamma_{ij}$ the spatial  3-metric.
The proper time $\tau$ is related to a coordinate time $t$ by
\be
\Theta d\tau = dt,
\label{eq:Lorentz_factor}
\ee
where a generalization of the Lorentz-factor
\be
\Theta\equiv \frac{1}{\sqrt{-g_{\mu\nu} v^\mu v^\nu}} \quad {\rm with} \quad v^\alpha=\frac{dx^\alpha}{dt}
\label{eq:theta_def}
\ee
was introduced. This relates to the four-velocity $U^\nu$, normalized to $U^\mu U_\mu= -1$, by
\be
v^\mu= \frac{dx^\mu}{dt}= \frac{dx^\mu}{d\tau} \frac{d\tau}{dt}= \frac{U^\mu}{\Theta}= \frac{U^\mu}{U^0}.
\label{eq:v_mu}
\ee
The Lagrangian of an ideal relativistic fluid  can be written as \cite{fock64}
\be
L= - \int T^{\mu \nu} U_\mu U_\nu \sqrt{-g} dV,
\label{eq:lag_ideal_fluid}
\ee
where $g= {\rm det}(g_{\mu \nu})$ and $T^{\mu\nu}$ denotes the energy-momentum tensor of an ideal
fluid without viscosity and conductivity
\be
T^{\mu \nu}= (\rho+P)U^\mu U^\nu + P g^{\mu \nu}.
\label{eq:Tmunu}
\ee
The local energy density (for clarity including the speed of light) is given by
\be
\rho= \rho_{\rm rest} + \frac{u \rho_{\rm rest}}{c^2}= n m_0 c^2 \left(1 + \frac{u}{c^2}\right).
\ee
Here $u$ is the specific internal energy per rest mass and $n$ the baryon number density as measured 
in the rest frame of the fluid. From now on, we follow the convention that 
all energies are measured in units of $m_0 c^2$, where $m_0$ is the baryon mass (and we use again $c= 1$).\\
The procedure to arrive at a set of SPH evolution equations is, as in the Newtonian and 
special-relativistic case, to first discretize the Lagrangian and  then apply the Euler-Lagrange
equations. In the relativistic case it is advantageous to use canonical momentum 
and energy (see Eqs.~(\ref{eq:can_mom}) and (\ref{eq:can_en}) below)
 as numerical variables, while in the Newtonian case one instead usually uses
a straight forward discretization of the first law of thermodynamics for the energy equation. Another peculiarity
of the relativistic case is that, due to Lorentz contractions, one has to carefully distinguish
between the local fluid rest frame (in which thermodynamic quantities are usually defined)
and the chosen ``computing frame'' in which the simulations are performed.\\
To find a SPH discretization in terms of a suitable density variable one can express
local baryon number conservation, $(U^\mu n);_\mu= 0$, as \cite{siegler00a}
\be
\p_\mu (\sqrt{-g} U^\mu n)= 0,
\ee
or, more explicitly, as
\be
\p_t (N) + \p_i(N v^i)= 0, \label{eq:continuity_N}
\ee
where Eq.~(\ref{eq:v_mu}) was used and the computing frame baryon number 
density\footnote{Note that the corresponding density, $\rho^\ast= \sqrt{-g} U^0 \rho$ is
also used in Eulerian formulations of Numerical Relativity, see e.g. \cite{etienne08} or \cite{shibata16}.}
\be
N= \sqrt{-g} \Theta n\label{eq:N_n}
\ee
was introduced.
The total conserved baryon number $\mathcal{N}$ can then be expressed as a sum over fluid parcels
with volume $\Delta V_b$ located at $\vec{r}_b$, where each fluid parcel carries a baryon number $\nu_b$
\be
\mathcal{N}= \int N dV \simeq \sum_b N_b \Delta V_b = \sum_b \nu_b \label{eq:parcel_volumes}
\ee
and $\Delta V_b= \nu_b/N_b$ is the volume assigned to particle $b$.  If we fix $\nu_b$ for each 
particle there is no need to solve a continuity equation (it can be done, though, if desired) and we can 
just calculate the computing frame number density at the position of a particle $a$ by 
\be
N_a= \sum_b \nu_b W(|\vec{r_a} - \vec{r}_b|,h_a),
\label{eq:N_sum}
\ee
where the smoothing length $h_a$ characterizes the support size of the smoothing kernel $W$.
Using the above, the Lagrangian of Eq.(\ref{eq:lag_ideal_fluid}) can now be straight forwardly discretized as
\be
L= - \sum_b \nu_b \left( \frac{1+u}{\Theta} \right)_b.
\ee
We use the canonical momentum per baryon of a particle $a$ as numerical variable
\be
(S_i)_a \equiv \frac{1}{\nu_a}\frac{\partial L}{\partial v^i_a} = (\Theta \mathcal{E} v_i)_a,
\label{eq:can_mom}
\ee
where $\mathcal{E}= 1 + u + P/n$ is the relativistic enthalpy per baryon and $v_i= g_{i \mu} v^\mu$,
we find the momentum evolution from the Euler-Lagrange equations as 
\be
\frac{d(S_i)_a}{dt}  =  \left(\frac{d(S_i)_a}{dt}\right)_{\rm hyd} +  \left(\frac{d(S_i)_a}{dt}\right)_{\rm met}
\label{eq:dSdt_full}
\ee
with
\be
\left(\frac{d(S_i)_a}{dt}\right)_{\rm hyd}= -\sum_b \nu_b \left\{ \frac{P_a}{N_a^2}  D^a_i  +  
\frac{P_b}{N_b^2} D^b_i \right\}
\label{eq:dSdt_hydro}
\ee
and
\be 
\left(\frac{d(S_i)_a}{dt}\right)_{\rm met}= \left(\frac{\sqrt{-g}}{2N} T^{\mu \nu} \frac{\p g_{\mu \nu}}{\p x^i}\right)_a.
\label{eq:dSdt_metric}
\ee
In the hydrodynamic terms we have used the convenient abbreviations
\be
D^a_i \equiv   \sqrt{-g_a} \;  \frac{\p W_{ab}(h_a)}{\p x_a^i} \quad {\rm and} \quad 
D^b_i \equiv    \sqrt{-g_b} \; \frac{\p W_{ab}(h_b)}{\p x_a^i}.
\ee
Starting from the canonical energy, $E= \sum_a (\partial L/\partial \vec{v}_a) \cdot \vec{v}_a -L$, one can define 
the canonical energy per baryon 
\be
e_a= \left(S_i v^i + \frac{1 + u}{\Theta}\right)_a = \left(\Theta \mathcal{E} v_i v^i + \frac{1 + u}{\Theta}\right)_a,
\label{eq:can_en}
\ee
which we use as numerical energy variable. Its evolution equation  follows\footnote{See \cite{rosswog09b}, Sec.~4.2, for the detailed steps.} from the differentiation of Eq.~(\ref{eq:can_en}) as
\be
\frac{d e_a}{dt}= \left(\frac{d e_a}{dt}\right)_{\rm hyd}  + \left(\frac{de_a}{dt}\right)_{\rm met},
\label{eq:energy_equation}
\ee
with
\be
\left(\frac{d e_a}{dt}\right)_{\rm hyd} = -\sum_b \nu_b \left\{ \frac{P_a}{N_a^2}  \;  v_b^i   \; D^a_i +  
\frac{P_b}{N_b^2} \;  v_a^i \; D^b_i \right\}
\label{eq:dedt_hydro}
\ee
and
\be
\left(\frac{de_a}{dt}\right)_{\rm met}= -\left(\frac{\sqrt{-g}}{2N} T^{\mu \nu} \frac{\p g_{\mu \nu}}{\p t}\right)_a.
\label{eq:dedt_metric}
\ee
With these momentum and energy variables the evolution equations are formally very similar
to the corresponding Newtonian equations. One important difference, however, is that the physical
(``primitive'') variables need to be reconstructed from the numerical (``conservative'') variables via
numerical root-finding techniques. How this is done in  \SpB is explained in detail in Sec.~\ref{sec:recovery}.

\subsubsection{SPH kernel}
\label{sec:SPH_kernel}
The SPH equations from Sec.~\ref{sec:ideal_hydro}  use a kernel function $W$ to
estimate the computing frame density, Eq.(\ref{eq:N_sum}), and to calculate the pressure gradient terms 
in Eqs.~(\ref{eq:dSdt_hydro}) and (\ref{eq:dedt_hydro}). We
have implemented a large set of different SPH kernel functions into our kernel module, but for 
all of the shown tests we employ the Wendland C6-smooth kernel \cite{wendland95}
\be
W(q)= \frac{\sigma}{h^3} (1-q)^8_+ (32 q^3 + 25 q^2 + 8 q + 1),
\ee
where the normalization $\sigma= 1365/(64 \pi)$ in 3D and the symbol $(.)_+$ denotes the cutoff function max$(.,0)$.  This kernel has provided excellent results 
in an extensive test series \cite{rosswog15b,rosswog20a}. It needs, however,  a large particle number in 
its support for good estimates of densities and gradients. Here we choose the smoothing length 
of each particle so that exactly 300 particles contribute.  To find the neighbour particles
we use a trimmed-down version of the tree-code described in detail in \cite{gafton11}. The technical
procedure is exactly the same as in the Newtonian SPH code \Ma and we refer to the corresponding code
paper \cite{rosswog20a} for a description of how this is done.
Similar to Liptai \& Price \cite{liptai19}, we use the Euclidian distance in Cartesian coordinates,
$d_{ab}= \sqrt{\eta_{ij} r^i_{ab} r^j_{ab}}$, as the
distance measure  that enters kernel evaluations such as $W(d_{ab},h_a)$. Here 
$r^j_{ab}= r^j_a - r^j_b$ is the difference between the contra-variant position vectors. 
For later use we also introduce $\hat{e}_{ab}= (r^j_{ab})/d_{ab}$.

\subsubsection{Dissipative terms}
The equations  in Sec.~\ref{sec:ideal_hydro} do not contain any way
to produce entropy and therefore they need to be enhanced by additional measures
to handle shocks. 
Entropy can be created either via  Riemann solvers or by applying artificial viscosity. Here we follow the latter 
approach, but we apply  techniques that are similar to those used in the context of
approximate Riemann solvers. We perform in particular a slope-limited reconstruction between 
particle pairs, a technique that has turned out to be a major improvement in Newtonian SPH \cite{rosswog20a}.
In their special-relativistic study \cite{chow97} suggested a dissipation scheme that did not distinguish
between artificial viscosity and conductivity. While able to robustly handle strong shocks, this scheme lead
to an excessive smoothing of contact discontinuities. In a recent analysis,  \cite{liptai19} suggested
a split between viscosity and conductivity. We follow a similar approach in this work, 
but we enhance their strategy by using slope-limited reconstructions and we steer 
the amount of dissipation by monitoring the entropy conservation, similar to what has been done
in a Newtonian context by \cite{rosswog20b}.\\

{\em Artificial viscosity} \\
Artificial viscosity can be easily implemented by simply adding an additional viscous contribution $Q$
to the physical pressures $P$, i.e. by replacing  $P$, wherever it occurs in the SPH equations, with
$P+Q$ \cite{vonneumann50}. We implement the viscous pressures suggested in \cite{liptai19}
\bea
Q_a&=& -\frac{1}{2} \alpha_{\rm AV} N_a v_{{\rm s},a} \enth_a \left(\Gamma_a^\ast V^\ast_a -  \Gamma_b^\ast V^\ast_b \right)
\label{eq:Qa}
\\
Q_b&=& -\frac{1}{2} \alpha_{\rm AV} N_b v_{{\rm s},b} \enth_b \left(\Gamma_a^\ast V^\ast_a -  \Gamma_b^\ast V^\ast_b \right),
\label{eq:Qb}
\eea
where the $V^\ast$ are the velocities of a Eulerian observer projected onto the line connecting particles $a$ and $b$,
\be
V^\ast_a= \eta_{ij} \hat{e}^j_{ab} V_a^i \quad{\rm and} \quad \Gamma_a^\ast= \frac{1}{\sqrt{1 - V_a^{\ast 2}}}
\ee
and correspondingly for $V^\ast_b$. The Eulerian observer velocity $V^i$ is related to the 
coordinate velocity $v^i$ by
\be
V^i= \frac{v^i + \beta^i}{\alpha}. 
\ee 
For the signal speeds we use
\be
v_{\rm s,a}= \frac{c_{\rm s,a} + |V^\ast_{ab}|}{1 + c_{\rm s,a}  |V^\ast_{ab}|},
\ee
where $c_{\rm s}= \sqrt{(\Gamma-1)(\mathcal{E}-1)/\mathcal{E}}$ is the relativistic sound speed and
\be
V^\ast_{ab}= \frac{V^\ast_a - V^\ast_b}{1 - V^\ast_a V^\ast_b}.
\ee


{\em Artificial conductivity}\\
To include artificial conductivity  we add the following term to our energy equation~(\ref{eq:energy_equation})
\be
\left(\frac{de}{dt}\right)^{\rm c}= \frac{\alpha_u}{2}  \sum_b \nu_b \xi^u_{ab} \left( \frac{\alpha_a u_a}{\Gamma_a} - \frac{\alpha_b u_b}{\Gamma_b}\right) \left\{ \frac{v_{\rm s, a}^u D^a_i }{N_a} +   \frac{v_{\rm s, b}^u D^b_i }{N_b}\right\} \hat{e}_{ab}^i,
\label{eq:cond}
\ee
where the $\alpha_a/\alpha_b$ are the lapse functions at the particle positions (not to be confused with $\alpha_{\rm AV}$)
and $ \Gamma= (1 - V_i V^i)^{-1/2}$. This conductivity term is, apart from the limiter $\xi^u_{ab}$ described below,
the same as in \cite{liptai19}. For the conductivity signal  velocity we use \cite{liptai19}
\be
v_{\rm s}^u= \rm{min} \left(1, \sqrt{\frac{2|P_a-P_b|}{\ent_a n_a + \ent_b n_b}}\right)
\ee
for cases when the metric is known (i.e. cases where no consistent hydrostatic equilibrium needs to be maintained)
and $v_{\rm s}^u= |V^\ast_{ab}|$ otherwise. For the prefactor $\alpha_u$ we chose after some experimenting a value of 0.3.\\
Conductivity can have detrimental effects if, for example, it spuriously switches on in a self-gravitating system
like a star. In such a case it can drive the star out of hydrostatic equilibrium.  In our applications we actually only want conductivity to act
where second derivatives, $\partial_i \partial_j u$ are large, for example near a contact discontinuity in a shock, otherwise
we want to suppress it. To this end we design a simple dimensionless trigger to measure the size of second-derivative effects
\begin{equation}
T_{u, ab}= \frac{h_{ab}}{u_{ab}} | (\nabla u)_a - (\nabla u)_b|,
\end{equation}
where $u_{ab}= (u_a+u_b)/2$ and $h_{ab}= (h_a+h_b)/2$. When this dimensionless quantity is large,
 we want conductivity to act, but otherwise it should be suppressed.
We achieved this by inserting the limiter
\begin{equation}
\xi^u_{ab}= \frac{T_{u, ab}}{T_{u, ab} + 0.01}
\end{equation}
inside the sum in Eq.~(\ref{eq:cond}), the reference value 0.01 has been chosen after experiments
in both Sod-type shock tubes and self-gravitating neutron stars.\\

{\em Reconstruction}\\
The above described artificial dissipation, Eqs.(\ref{eq:Qa}) and (\ref{eq:Qb}),  contains
``jumps'' of quantities measured at the particle positions. In the 
lingo of Finite Volume Methods (FVM)  this is called a ``zeroth order reconstruction''. In FVM one 
usually ``reconstructs''  fluid variables from the cell centres to the interfaces between two 
adjacent cells and there one applies (exact or approximate) Riemann solver techniques 
to these reconstructed variables  to obtain the numerical fluxes between the cells. Increasing 
the polynomial order of the reconstruction usually reduces the diffusivity of a numerical scheme. 
In the reconstruction process one usually applies  ``slope limiters'' to the original 
gradient estimates to avoid introducing new maxima or minima.\\
Although we neither use a FVM nor solve a Riemann problem, the above described 
techniques can nevertheless be applied to our artificial dissipation scheme: instead of using the differences
of the quantities at the particle positions, we use the differences between the reconstructed quantities
at the inter-particle position. In  Newtonian hydrodynamics \cite{frontiere17,rosswog20a} such an 
approach was found to drastically reduce the net dissipation, even when constant large dissipation 
parameters were used.\\
Consider two particles $a$ and $b$ with (contra-variant) position vectors $r^i_a$ and $r^i_b$.
For the artificial pressures, we reconstruct the Eulerian observer velocity from the $a$-side of
the mid-point between the particles, $r^i_{ab}= (r^i_a + r^i_b)/2$, as
\be
\tilde{V}^i_a= V^i_a - \frac{1}{2} {\rm SL}(\partial_jV^i_a,\partial_jV^i_b) (r^j_a - r^j_b),
\label{eq:recon_a}
\ee
the corresponding velocity from the $b$-side reads
\be
\tilde{V}^i_b= V^i_b + \frac{1}{2} {\rm SL}(\partial_jV^i_a,\partial_jV^i_b) (r^j_a - r^j_b).
\label{eq:recon_b}
\ee
We experimented with several standard slope-limiter functions SL: minmod, vanLeer, 
vanLeerMC \cite{vanLeer74,vanLeer77} and superbee \cite{roe86}. While many combinations give good
results, we usually need higher dissipation parameters when using less-dissipative limiters. Therefore
we have settled on the simplest (most dissipative and robust) limiter minmod,
\be
{\rm SL}^{\rm minmod} (a,b)=    \left\{
    \begin{array}{ll}
    \; \; \;{\rm min}(|a|,|b|) & \rm{if \; }  a> 0 {\rm \; and \;} b > 0\\
   -  {\rm min}(|a|,|b|) & {\rm if \;} a < 0 {\rm \; and \;} b < 0\\    
    \; \; \;0 & {\rm otherwise,}
   \end{array}.
 \right.
 \ee
together with moderate values for the dissipation parameters, see below.
In our artificial viscosity scheme with reconstruction
we  apply the artificial pressures as described above in Eq.(\ref{eq:Qa}) and (\ref{eq:Qb}), but we calculate them
using $\tilde{V}^i_a$ and $\tilde{V}^i_b$ instead of  $V^i_a$ and $V^i_b$. \\
We proceed similarly for the conductive terms where we use reconstructed values $\tilde{u}_a$ and $\tilde{u}_b$
in Eq.(\ref{eq:cond}) instead of $u_a$ and $u_b$, where the reconstructed values are obtained analogously to Eqs.~(\ref{eq:recon_a})
and (\ref{eq:recon_b}). We will illustrate the beneficial effects of reconstruction in the context of a shock test, see 
Fig.~\ref{fig:Sod_recon}.\\

{\em Steering dissipation via entropy conservation}\\
While the reconstruction already dramatically reduces the unwanted effects of excessive dissipation \cite{rosswog20a},
one can actually even go one step further and also make the dissipation parameter $\alpha_{\rm AV}$ time dependent. We implement
here the dissipation steering strategy suggested in \cite{rosswog20b}. The main idea is that an ideal fluid should --in the absence of shocks--
conserve entropy exactly. If shocks are present, they can increase the entropy, but entropy violations can also occur
for purely numerical reasons, if, for example, the flow becomes ``noisy'' with substantial velocity fluctuations. In both cases
one wants to add dissipation (to either resolve the shocks properly or calm down the noisy flow) and we therefore use
entropy conservation violations as a measure to identify ``troubled particles'' and to assign to each particle a desired 
dissipation parameter value, $\alpha_{{\rm AV}, a}^{\rm des}$. If this value is larger than the current value $\alpha_{{\rm AV}, a}(t)$,
the latter is instantly increased to $\alpha_{{\rm AV}, a}^{\rm des}$. Otherwise, the dissipation parameter  decays according to
\be
\frac{d\alpha_{{\rm AV}, a}}{dt}=  -\frac{\alpha_{{\rm AV}, a}(t) - \alpha_0}{ \tau_a},
\label{eq:davdt}
\ee
where for the decay time scale we use $\tau_a= 30 h_a/c_{\rm s,a}$. What remains is to assign a value of $\alpha_{{\rm AV}, a}^{\rm des}$
based on the entropy violations. To this end we monitor the logarithm of the relative entropy change at each particle between two time steps
\be
l^n_a\equiv \log_{10} \left(\frac{K^n_a - K^{n-1}_a}{K^{n-1}_a} \right),
\ee
where the index $n$ indicates a value at time $t^n$, $K_a$ is ``pseudo-entropy'' $K_a= P_a/n_a^\Gamma$  and $\Gamma$ the
polytropic exponent.
If $l^n$ is below an acceptable threshold value, $l_0=-5$, $\alpha_{{\rm AV}, a}^{\rm des}= \alpha_0$, if it is
above a value where we want full dissipation, $l_1=-2$, we set $\alpha_{{\rm AV}, a}^{\rm des}= \alpha_{\rm AV}^{\rm max}$,
and in between the desired value is calculated via
\be
\alpha_{a,\rm AV}^{\rm des}= (\alpha_{\rm max} - \alpha_0 )\; \mathcal{S}(l_a^n) + \alpha_0
\ee
with the smooth switch-on function
\be
\mathcal{S}(x)= 6x^5 - 15x^4 + 10x^3
\ee
and
\be
x= \rm min\left[max\left(\frac{l_a^n - l_0}{l_1-l_0},0\right),1\right].
\ee
For the shape of the switch-on function we refer to Fig.~1 in the original paper \cite{rosswog20b}.
As our default parameters we choose $\alpha_0= 0.1$  and $\alpha_{\rm max}= 1.5$.

\subsubsection{Recovery of primitive variables}
\label{sec:recovery}
As in Eulerian relativistic hydrodynamics, we need to recover the physical (``primitive'') variables $u, n, v^i$ from 
the numerical (``conservative'') ones  $N, S_i, e$, see Eqs.~(\ref{eq:N_sum}),   
(\ref{eq:can_mom}) and  (\ref{eq:can_en}). For now, we restrict ourselves to a polytropic equation
of state which, with our conventions, reads
\be
P= (\Gamma-1) n u.
\label{eq:poly_EOS}
\ee 
The strategy is to express $n$ and $u$ in terms of the known numerical variables $N, S_i, e$ and the pressure
$P$, substitute these expressions in Eq.(\ref{eq:poly_EOS}) and solve the resulting equation
\be
f(P) \equiv P - (\Gamma-1) \; n(S_i,e,P) \; u(S_i,e,P)= 0,
\label{eq:recovery_to_solve}
\ee
for a new, consistent value of $P$. Once this value is found, the primitive variables are  recovered by
back-substituting the new values of $N, S_i, e$ and $P$.\\
We start by solving $1= dt/dt= v^0= g^{0\mu}v_\mu$ for 
\be
v_0= \frac{1-g^{0i}S_i/(\Theta \mathcal{E})}{g^{00}},
\label{eq:dum1}
\ee
which can be used to solve Eq.~(\ref{eq:theta_def}) for $v_i v^i$. The latter can be used in Eq.~(\ref{eq:can_en})
to find
\be
e= \frac{g^{0j}S_j}{g^{00}} - \frac{P}{\Theta n} - \frac{\Theta \mathcal{E}}{g^{00}},
\label{eq:e_v1}
\ee
which we solve for the internal energy (as expressed in the desired variables)
\be
u= \frac{g^{0j} S_j}{\Theta} - \frac{g^{00} e}{\Theta} - \frac{\sqrt{-g}P}{\Theta N} \left( g^{00} + \Theta^2\right) - 1.
\label{eq:u}
\ee
Using Eq.(\ref{eq:N_n}) we solve the equation for the canonical energy, Eq.~(\ref{eq:e_v1}) for $\Theta \ent$,
which, in turn, provides the co-variant velocity components 
\be
v_i= \frac{S_i}{\Theta \mathcal{E}}= S_i \left[ g^{0j} S_j - g^{00} \left( \frac{\sqrt{-g} P}{N} + e \right) \right]^{-1}\label{eq:vi}
\ee
from Eq.~(\ref{eq:can_mom}).  The generalized Lorentz factor can be expressed as
\be
\Theta= \sqrt{\frac{-g^{00}}{1 + \frac{A}{B^2}}}
\label{eq:theta1}
\ee
where
\be
A = g^{00} g^{jk}S_j S_k - (g^{0j} S_j)^2\\
\ee
and
\be
B = g^{0j} S_j - g^{00} \left(\frac{\sqrt{-g}}{N} P + e \right).
\ee
Using Eq.~(\ref{eq:theta1}) and (\ref{eq:N_n}) we find $n$ which can, together with Eq.~(\ref{eq:u}), be inserted into
Eq.~(\ref{eq:recovery_to_solve}) to find the new, consistent pressure value  $P$ by means of  a Newton-Raphson scheme.
The desired primitive variables are then found by back-substitution: $\Theta$ from Eq.~(\ref{eq:theta1}), $v_i$ from
(\ref{eq:vi}), $n$ from (\ref{eq:N_n}) and the internal energy $u$ from Eq.~(\ref{eq:u}).

\subsection{Spacetime evolution}
\label{sec:spacetime_evolution}
In \spB, we have two of the frequently used variants of the BSSN equations implemented, 
the so-called ``$\Phi$-'' and  the ``$W$-method. We extracted  the code for
these from the \Mcl thorn~\cite{Brown:2008sb} in the Einstein 
Toolkit~\cite{ETK:web,Loffler:2011ay} and
build our own wrapper function to call all the needed functions. This was done
partially in order to not have to, yet again, reimplement the BSSN equations
and partially to start out with a well tested implementation.

As our default, we use the so-called  ``$\Phi$-method'' \cite{shibata95,baumgarte99}, the variables of which are based on the 
ADM variables $\gamma_{ij}$ (3-metric), $K_{ij}$ (extrinsic curvature),
$\alpha$ (lapse) and $\beta^{i}$ (shift) and they read
\begin{eqnarray}
  \phi & = & \frac{1}{12} \log(\gamma), \label{eq:bssn_phi}\\
  \tlg_{ij} & = & \emfp \gamma_{ij}, \\
  K & = & \gamma^{ij} K_{ij}, \\
  \tlG^{i} & = & \tlg^{jk} \tlG^{i}_{jk}, \\
  \tlA_{ij} & = & \emfp\left ( K_{ij}-\frac{1}{3}\gamma_{ij} K\right )\label{eq:bssn_A},
\end{eqnarray}
where $\gamma = \mathrm{det}(\gamma_{ij})$,  $\tlG^{i}_{jk}$ are the 
Christoffel symbols related to the conformal metric $\tlg_{ij}$ and $ \tlA_{ij}$ is
the conformally rescaled, traceless part of the extrinsic curvature. The corresponding
evolution equations read
\begin{eqnarray}
  \dt{\phi} & = & -\frac{1}{6} \left ( \alpha K - \pdu{\beta}{i}{i} \right) + \upwindu{\phi}{}{i}, \label{eq:bssn_ev_start}\\
  \dt{\tlg_{ij}} & = & -2\alpha \tlA_{ij} + \tlg_{ik} \pdu{\beta}{k}{j}
                   + \tlg_{jk} \pdu{\beta}{k}{i}
                    -\frac{2}{3} \tlg_{ij} \pdu{\beta}{k}{k}\nonumber \\
                       & &  + \upwindl{\tlg}{ij}{k}, \\
  \dt{K} & = & -\emfp \left ( \tlg^{ij} \left [ \pdpdu{\alpha}{}{i}{j}
               +2\pdu{\phi}{}{i}\pdu{\alpha}{}{j} \right ] 
               - \tlGn^{i}\pdu{\alpha}{}{i} \right ) \nonumber \\
         &   & + \alpha \left ( \tlA^{i}_{j} \tlA^{j}_{i} +\frac{1}{3} K^2
               \right ) + \upwindu{K}{}{i} + 4 \pi \alpha ( \rho + s ), \\
  \dt{\tlG^{i}} & = & -2 \tlA^{ij} \pdu{\alpha}{}{j} + 2 \alpha \left (
                    \tlG^{i}_{jk} \tlA^{jk} - \frac{2}{3} \tlg^{ij}
                    \pdu{K}{}{j} +  
                  6 \tlA^{ij} \pdu{\phi}{}{j}\right )
                 \nonumber\\
                  & &                     +\tlg^{jk} \pdpdu{\beta}{i}{j}{k} + \frac{1}{3}
                    \tlg^{ij} \pdpdu{\beta}{k}{j}{k} -\tlGn^{j}\pdu{\beta}{i}{j}
                    + \frac{2}{3} \tlGn^{i}\pdu{\beta}{j}{j} \nonumber \\
                &   & + \upwindu{\tlG}{i}{j} -16 \pi \alpha \tlg^{ij} s_j, \\
  \dt{\tlA_{ij}} & = & \emfp \left [ -\pdpdu{\alpha}{}{i}{j} + \tlG^{k}_{ij}
                       \pdu{\alpha}{}{k} + 2 \left ( \pdu{\alpha}{}{i}
                       \pdu{\phi}{}{j}+\pdu{\alpha}{}{j} \pdu{\phi}{}{i}\right ) 
                       +\alpha R_{ij} \right ]^{\mathrm{TF}} \nonumber\\
                 & & +\alpha ( K \tlA_{ij}- 2 \tlA_{ik} \tlA^{k}_{j} )
                       + \tlA_{ik} \pdu{\beta}{k}{j} 
                       + \tlA_{jk} \pdu{\beta}{k}{i}
                       - \frac{2}{3} \tlA_{ij} \pdu{\beta}{k}{k} \nonumber \\
                 &   & +\upwindl{\tlA}{ij}{k} - \emfp \alpha 8 \pi
                       \left (T_{ij}-\frac{1}{3} \gamma_{ij} s\right ), 
		       \label{eq:bssn_ev_end}
                       \end{eqnarray}
where 
\begin{eqnarray}
  \rho & = & \frac{1}{\alpha^2} ( T_{00} - 2 \beta^{i} T_{0i} +
             \beta^{i}\beta^{j} T_{ij} ),\label{eq:BSSN_rho} \\
  s & = & \gamma^{ij} T_{ij}, \\
  s_{i} & = & -\frac{1}{\alpha} ( T_{0i} - \beta^{j} T_{ij}),\label{eq:BSSN_Si}
\end{eqnarray}
and $\upwindu{}{}{i}$ denote partial derivatives that are upwinded based on the
shift vector. Finally $R_{ij} = \tlR_{ij} + R^{\phi}_{ij}$ where
\begin{eqnarray}
  \tlG_{ijk} & = & \frac{1}{2}\left ( \pdl{\tlg}{ij}{k} + \pdl{\tlg}{ik}{j} 
               - \pdl{\tlg}{jk}{i} \right ), \\
  \tlGmixed{ij}{k} & = & \tlg^{kl} \tlG_{ijl}, \\
  \tlG^{i}_{jk} & = & \tlg^{il}\tlG_{ljk}, \\
  \tlGn^{i} & = & \tlg^{jk} \tlG^{i}_{jk} \\
  \tlR_{ij} & = & -\frac{1}{2} \tlg^{kl} \pdpdl{\tlg}{ij}{k}{l}
                  +\tlg_{k(i} \pdu{\tlG}{k}{j)}
                  +\tlGn^{k} \tlG_{(ij)k} \nonumber \\
            &   & +\tlG^{k}_{il} \tlGmixed{jk}{l}
                  +\tlG^{k}_{jl} \tlGmixed{ik}{l}
                  +\tlG^{k}_{il} \tlGmixed{kj}{l}, \\
  R^{\phi}_{ij} & = & -2\left (\pdpdu{\phi}{}{i}{j}
                 -\tlG^{k}_{ij}\pdu{\phi}{}{k}\right )
                 -2\tlg_{ij} \tlg^{kl} \nonumber\\
            & &     \left ( \pdpdu{\phi}{}{k}{l}
                 -\tlG^{m}_{kl}\pdu{\phi}{}{m}\right ) 
                + 4\pdu{\phi}{}{i}\pdu{\phi}{}{j} \nonumber\\
             & &    - 4\tlg_{ij}\tlg^{kl}\pdu{\phi}{}{k}\pdu{\phi}{}{l}.
\end{eqnarray}

For the gauge choices we use a variant of ``1+log''-slicing, where the lapse
is evolved according
to
\be
\partial_t \alpha= -2 \alpha K
\ee
and a variant of the ``gamma-driver'' shift evolution with
\be
\partial_t \beta^i= \frac{3}{4}(\tilde{\Gamma}^i-\beta^i).
\ee
\SpB still supports all the gauge choices implemented in \mcl, but we found
that these simple choices were sufficient for the simulations in this paper.
The derivatives are calculated via finite differencing of 4th, 6th or 8th
order. Unless mentioned otherwise, we use our fourth order finite
differencing as default.\\
We can of course not evaluate the evolution equations near the boundary
of the domain as the finite differencing stencils would require values from
grid points outside of the domain. Instead, we apply the same Sommerfeld-type
radiative boundary conditions as used in the Einstein Toolkit, see 
section 5.4.2 in \cite{Loffler:2011ay}, to all the evolved BSSN variables.

From the BSSN variables, the lapse and the shift, the physical 4-metric can
be reconstructed as
\be
g_{\mu\nu} = \pmatrix{
  -\alpha^2 +e^{4\phi}\tlg_{ij}\beta^{i}\beta^{j} & e^{4\phi}\tlg_{ik}\beta^k \cr
e^{4\phi}\tlg_{jk}\beta^k & e^{4\phi}\tlg_{ij}
  } .\label{eq:phys_metric}
\ee

\noindent In addition to the ``$\phi$-method'' we have also implemented the so-called ``$W$-method'' \cite{marronetti08,tichy07}, which we summarize
for completeness in Appendix A.\\

\subsection{Coupling the hydrodynamic and the spacetime evolution: a particle-mesh approach}
\label{sec:particle_mesh}
A crucial ingredient of our method is the interaction of the fluid (represented by particles) with the 
spacetime (represented on a mesh): the spacetime evolution needs the energy momentum-tensor, 
Eq.(\ref{eq:Tmunu}), at the grid points as an input, while the fluid  needs the metric and its derivatives, 
$g_{\mu\nu}$ and $\partial_\lambda g_{\mu \nu}$, at the particle positions for the evolution equations (\ref{eq:dSdt_metric}) 
and (\ref{eq:dedt_metric}).  During the time-integration we therefore have to, at every sub-step, map the particles (more precisely their
energy momentum tensor) to the grid (``P2M-step'') and grid properties (more precisely the metric and derivatives) back to
the particle positions (``M2P-step''). Similar steps are needed in other particle-mesh methods e.g.
in plasma physics simulations \cite{hockney88} or in vortex methods \cite{cottet00} and we draw some inspiration from
them.\\

\begin{figure}
   \centering
   \includegraphics[width=1.\columnwidth]{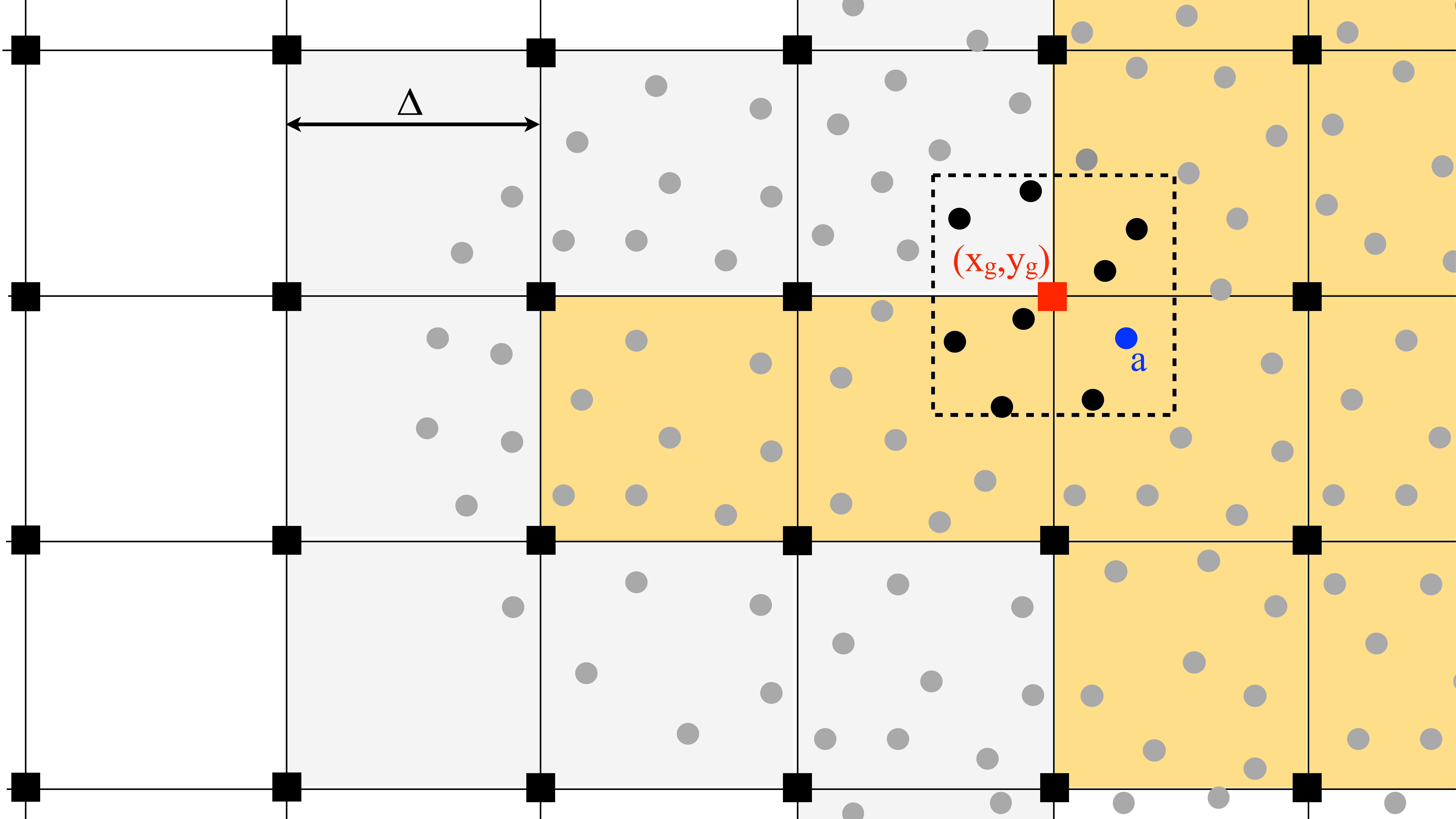} 
   \caption{Sketch of a particle mesh configuration (for simplicity in 2D). Surface cells (at least one empty direct neighbour cell)
   are underlaid with grey, cells with all direct neighbour cells being non-empty, but with at least one empty next-to-direct neighbour
   cell are underlaid with orange. The volume assigned to the grid point at $(x_g,y_g)$ is indicated by the dashed square.}
\label{fig:sketch_particle_mesh}
\end{figure}

\noindent{\em Preparation step}\\
We are, for simplicity, using a uniform Cartesian mesh with a mesh size $\Delta$.
As a first step we assign the particles to their closest grid point at $\vec{r}_{g}= (x_g,y_g,z_g)$, 
so that each grid point has a list of particles contained within 
$[x_g - \Delta/2, x_g + \Delta/2 ) \times [y_g - \Delta/2, y_g + \Delta/2 ) \times [z_g - \Delta/2, z_g + \Delta/2 )$.
In a second step, each cell is flagged according to the ``filling status'' ($fs$) of its neighbour cells, which will
later help to decide which mapping method to use. Filled (=non-empty) cells, which have at least the closest three neighbour cells
in each direction filled, receive label  $fs= 3$, cells with two filled neighbour cells in each direction are labelled with $fs= 2$
and so on. This is sketched for a 2D version in Fig.~\ref{fig:sketch_particle_mesh}.\\

\noindent{\em Kernel choice}\\
In order to map particle properties to the grid and back we use kernel techniques.  
To avoid potential confusion with the SPH-kernels, $W$, we refer to these ``shape functions'' as $\Psi$.
In SPH one usually chooses radial shape functions $W(\vec{r}-\vec{r}_b,h)= W(|\vec{r}-\vec{r}_b|,h)$ since this
allows, in a straight forward way, for exact conservation of angular momentum, see e.g. Sec. 2.4 in \cite{rosswog09b}
for a detailed discussion of conservation in SPH. Since the density is (most often and also here) calculated
as a kernel-weighted sum over nearby particles, see Eq.~(\ref{eq:N_sum}), one wants to use positive definite
kernels so that a positive density estimate is guaranteed under all circumstances.\\
We distinguish between the {\em degree} of the kernel (=degree of polynomial order), its (approximation) {\em order}
and its {\em regularity} (= number of times the kernel is continuously differentiable). 
While their positivity makes SPH kernels robust density estimators, it also limits them  
to (only) second order. Higher order interpolation  kernels have negative values
in parts of their support  and are therefore avoided in SPH \cite{monaghan92}. For the mapping
of particles to a mesh, however, such kernels can deliver accurate results, provided that they are not 
applied across sharp edges like the surface of the neutron star. If the latter happens, this leads to
disastrous oscillations that can result in unphysical values and code crashes. This is why 
we have assigned each cell a filling status flag which is used to decide which shape function to use.\\

\noindent{\em Particle-to-Mesh (P2M) step}\\

\begin{figure}
   \centering
   \includegraphics[width=0.8\columnwidth,angle=-90]{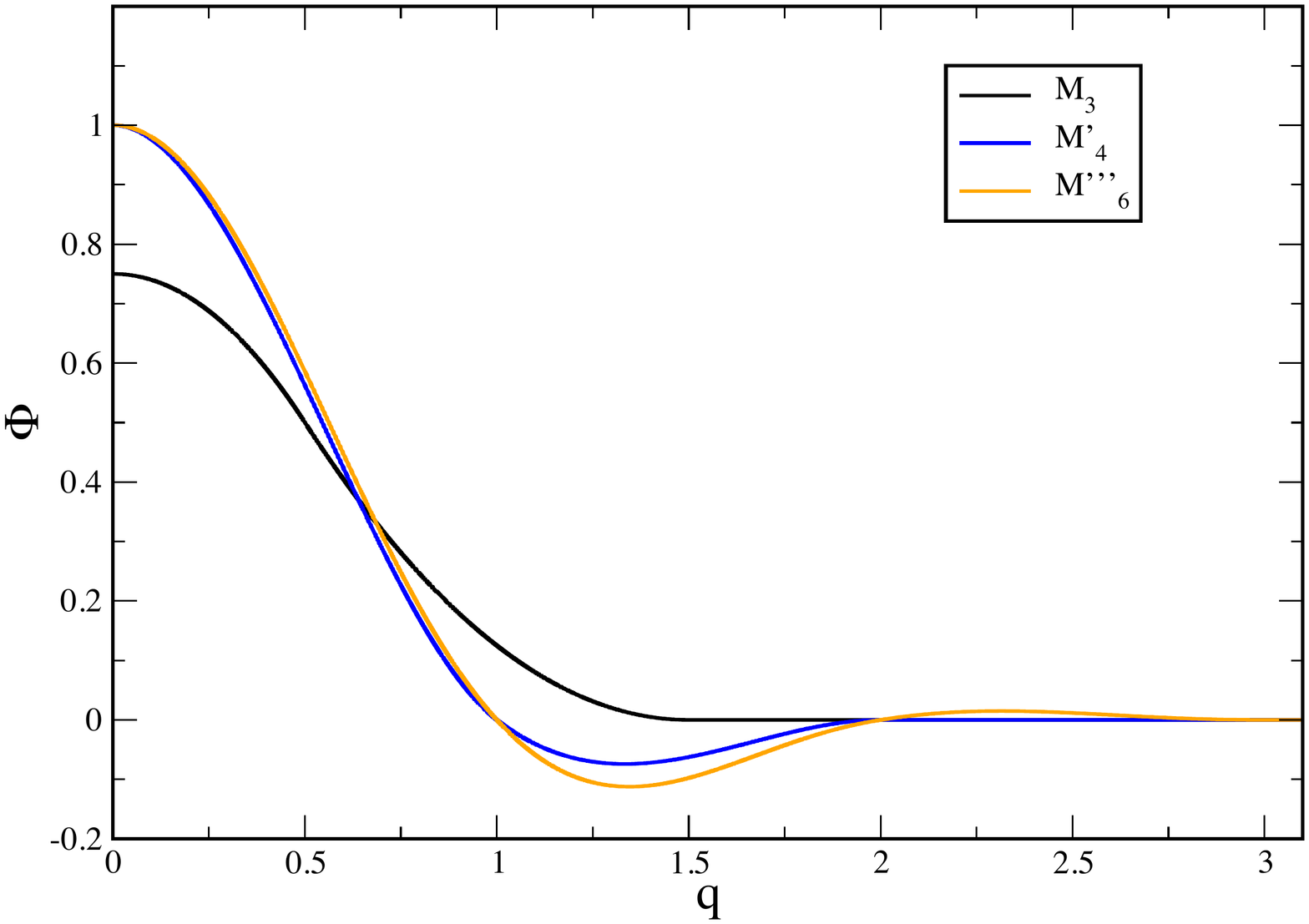} 
   \caption{The shape functions used in the ``particle-to-mesh'' mapping step.}
\label{fig:mapping_kernels}
\end{figure}

\noindent A. Pre-described shape functions\\
The P2M-step is the more challenging of both steps since the particles are not guaranteed to be regularly distributed
in space. Hence it is not straight forward to accurately assign their properties (here $T_{\mu\nu}$) to the 
surrounding grid points.
We map a quantity $A$ that is known at particle positions $r^i_p$ to the grid point  $r^i_g$
via 
\be
A_g= A(\vec{r}_g)=  \frac{\sum_p V_p A_p \Psi_g(\vec{r}_p)}{\sum_p V_p  \Psi_g(\vec{r}_p)},
\ee
where $V_p= \nu_p/N_p$ is a measure of the particle volume.
We apply here  a hierarchy of shape functions $\Psi$ of decreasing interpolation order depending
on the filling status of the neighbouring cells. In all of the cases we use tensor products of
1D functions
\be
\Psi(x,y,z)_g= \Phi\left(\frac{|x-x_g|}{\Delta}\right) \Phi\left(\frac{|y-y_g|}{\Delta}\right) \Phi\left(\frac{|z-z_g|}{\Delta}\right).
\label{eq:T_product}
\ee
We have experimented with a number of different shape functions, starting from commonly used SPH kernels,
each time monitoring how close a (low resolution) neutron star remains to its initial TOV solution when both the fluid and the
metric are evolved (typically monitoring several dozen dynamical time scales). We find good results for
the following hierarchy of 1D shape functions $\Phi$ (to be used in Eq.(\ref{eq:T_product})):
\begin{itemize}
	\item for cells with $fs=3$ and $fs=2$ we use \cite{bergdorf06}    
    \[
   \hspace*{-0.2cm} M^{'''}_6 (q)  \hspace*{-0.1cm} =  \hspace*{-0.1cm} \left\{\begin{array}{lr}
\hspace*{-0.2cm}         -\frac{1}{88}(q - 1)\left[60 q^4 - 87(q^3 + q^2) + 88(q + 1)\right]&  \hspace*{-0.2cm} q <  1\\
\hspace*{-0.2cm}          \frac{1}{176}  (q-1) (q-2)\left[60q^3 - 261 q^2 +257 q  + 68\right] &  \hspace*{-0.2cm} 1\leq q <2\\
\hspace*{-0.2cm}          -\frac{3}{176}  (q-2) \left[4q^2 - 17 q + 12 \right] (q-3)^2&  \hspace*{-0.2cm} 2\leq q < 3\\
        0 &\hspace*{-0.2cm} {\rm else},
        \end{array}\right. 
  \]
  \item for cells with  $fs=1$ we use \cite{monaghan85b,cottet00}  
   \[
   \hspace*{-0.2cm} M^{'}_4 (q)  \hspace*{-0.1cm} =  \hspace*{-0.1cm} \left\{\begin{array}{lr}
        1 - \frac{5}{2}q^2 + \frac{3}{2}q^3 &  \hspace*{-0.2cm} q <  1\\
        \frac{1}{2}(2-q)^2 (1-q)  &  \hspace*{-0.2cm} 1\leq q <2\\
        0 &\hspace*{-0.2cm} {\rm else}
        \end{array}\right. 
  \]
  \item and \cite{hockney88}
   \[
   \hspace*{-0.2cm} M_3(q)  \hspace*{-0.1cm} =  \hspace*{-0.1cm} \left\{\begin{array}{lr}
        \frac{1}{2}(q+\frac{3}{2})^2 - \frac{3}{2}(q+\frac{1}{2})^2 &  \hspace*{-0.2cm} q <  1/2\\
        \frac{1}{2}(-q+\frac{3}{2})^2  &  \hspace*{-0.2cm} 1/2 \leq q <3/2\\
        0 &\hspace*{-0.2cm} {\rm else}
        \end{array}\right. 
  \]  for cells with $fs=0$, i.e. for cells near the surface.
  \end{itemize}
Note that, strictly speaking,  with these choices the kernel support size can reach  empty cells beyond 
a fluid surface, but in all of our tests we found good results with the chosen hierarchy.  The kernels are plotted in 
Fig.~\ref{fig:mapping_kernels}. Note that out of these kernels, only $M_3$ is strictly positive definite.\\

%
%
\noindent B. Moving Least Squares\\
As an alternative to using the above described method with pre-described kernels, we have also implemented
a {\em Moving Least Squares} (MLS) approach to map the particle properties onto the mesh.
The main idea is to assign a set of basis functions $\{b^g_i(x,y,z), i=1..m\}$ to each grid point labelled by $g$ and to determine the 
needed set of coefficients $\{c^g_i, i=1..m\}$ by minimizing an error functional based on the particles
in the neighbourhood of the grid point.\\
The function to be mapped to the mesh, optimized at a grid point, is then
written as
\be
\tilde{A}_g(\vec{r})= \sum_{i=1}^m c^g_i b_i(\vec{r}).
\ee
The local coefficients $c^g_i$ are  determined by minimizing the error
functional
\be
\mathcal{L}(\{c^g_i\})\equiv \sum_b W(|\vec{r}_g - \vec{r}_b|) \left\{A_b - \sum_{i=1}^m c^g_i b_i(\vec{r}_b) \right\}^2
\ee
with respect to the $c^g_i$. The function $W$ gives more weight to nearby than to far away particles and one can take, for example, 
a typical SPH-kernel. Requiring
\be
0 \stackrel{!}{=} \frac{\partial \mathcal{L}}{\partial c^g_i}
\ee
yields the coefficients as
\be
c^g_i= M_{ij}^{-1} \; d_j,
\ee
where
\be
M_{ij}= \sum_b W(|\vec{r}_b-\vec{r}_g|) \; b_i(\vec{r}_b) \; b_j(\vec{r}_b) 
\ee
and 
\be
d_j= \sum_b= W(|\vec{r}_b-\vec{r}_g|) \; A_b \; b_j(\vec{r}_b).
\ee
In our approach we have chosen the basis functions 
$\{1, \tilde{x},\tilde{y},\tilde{z},\tilde{x}\tilde{x},\tilde{x}\tilde{y}, \tilde{x}\tilde{z},\tilde{y}\tilde{y},\tilde{y}\tilde{z},\tilde{z}\tilde{z}\}$,
where $\tilde{x}= x- x_g, \tilde{y}= y - y_g, \tilde{z}= z- z_g$, and a tensor-product version of the $M_3$ kernel
as  the positive-definite weight function. The required solution of a
$10\times10$ linear system involving the matrix  $(M_{ij})$ is performed via a LU-decomposition (and a singular
value decomposition \cite{press92} as fallback option) and this 
makes the  MLS approach for the P2M-step about 10\% more computationally expensive than
the prescribed kernels, but  in terms of the overall run time both approaches are very similar.\\

\noindent{\em Mesh-to-Particle (M2P) step}\\
Due to the regularity of a mesh, this step is somewhat simpler and we can draw 
on knowledge form mesh-based methods. An obvious choice would be to use exactly the
same kernels as in the P2M-step. After many numerical experiments, we have settled, however,
on two other methods, a WENO5-variant \cite{kozak20} and a quintic Hermite polynomial interpolation
that are substantially more accurate; in particular near the stellar surface. In the following, we will concisely
summarize these methods.\\

\noindent{A. WENO 5}\\
When interpolating some function, $A_g$, given at grid positions $\vec{r}_g$, to some general position
$\vec{r}$, oscillations can occur when encountering sharp transitions. Whether they occur or not depends
on the chosen stencil, and {\em Weighted Essentially Non-Oscillatory} (WENO) schemes are designed so
that a suitably weighted superposition of stencils gives most weight to non-oscillatory stencils. Here we follow
the suggestion of Kozak et al. \cite{kozak20} for such a scheme of fifth order (WENO5).\\
The task is now to ``transfer''  a function  that is known on a grid ($A_g$) to a general position
\begin{equation}
A(\vec{r})= \sum_g \Phi_g(\vec{r}) A_g,
\end{equation}
where the weight functions $\Phi_g$ form a partition of unity
\be
\sum_g \Phi_g(\vec{r})= 1.
\ee
Here, we also use tensor products of 1D-functions similar to Eq.~(\ref{eq:T_product}).
The scheme uses non-dimensional distances from the grid centres
\be
\tilde{x}= \frac{x-x_g}{\Delta} \quad \tilde{y}= \frac{y-y_g}{\Delta} \quad  \tilde{z}= \frac{z-z_g}{\Delta} 
\ee
and the following linear weights for the left, central and right positions
\bea
C^L&=& \frac{1}{12}(\xt -1)(\tilde{x}-2); \nonumber\\
C^C&=& -\frac{1}{6}(\xt + 2)(\xt-2);  \\
C^R&=& \frac{1}{12}(\xt+2)(\xt+1). \nonumber
\eea
The following smoothness indicators are used
\bea
\beta^L_{j,k}&=&\frac{13}{12} (A_{0,j,k} - 2 A_{1,j,k} + A_{2,j,k})^2 + \frac{1}{4} (A_{0,j,k} - 4 A_{1,j,k}  + 3 A_{2,j,k})^2\nonumber \\
\beta^C_{j,k}&=&\frac{13}{12}  (A_{1,j,k} - 2 A_{2,j,k} + A_{3,j,k})^2 + \frac{1}{4} (A_{1,j,k} - A_{3,j,k})^2  \\
\beta^R_{j,k}&=&\frac{13}{12} (A_{2,j,k} - 2 A_{3,j,k} + A_{4,j,k})^2 + \frac{1}{4} (3 A_{2,j,k} - 4 A_{3,j,k}  + A_{4,j,k})^2 \nonumber
\eea
and from them  the auxiliary variables 
\be
\alpha^M_{j,k}= \frac{C^M}{(\beta^M_{j,k} + \epsilon)^2}
\ee
are calculated, where $M$ stands for either $L,C$ or $R$. These $\alpha^M_{j,k}$ are then in turn used
for the non-linear weights
\be
\omega^M_{j,k}= \frac{\alpha^M_{j,k}}{\sum_I \alpha^I_{j,k}},
\ee
where the $I$ summation runs over $L,C$ and $R$. The final weight function is then
 \[
   \hspace*{-0.2cm} \Phi^{\rm W5}(\tilde{x})  \hspace*{-0.1cm} =  \hspace*{-0.1cm} \left\{\begin{array}{lr}
        0 &  \hspace*{-0.2cm} \xt <  -\frac{5}{2}\\
        \frac{1}{2}(\xt + 1)\; \xt \; \omega^L_{j,k} &  \hspace*{-0.2cm} -\frac{5}{2} \leq \xt <-\frac{3}{2}\\
        -(\xt+2) \; \xt \; \omega^L_{j,k} + \frac{1}{2}\xt \; (\xt -1) \; \omega^C_{j,k}&  \hspace*{-0.2cm} -\frac{3}{2}\leq \xt < -\frac{1}{2}\\
         \frac{1}{2}(\xt+2) (\xt+1) \;  \omega^L_{j,k} \\
	 -(\xt+1) (\xt-1)\; \omega^C_{j,k} &  -\frac{1}{2}\leq \xt < \frac{1}{2}\\
	 +\frac{1}{2}(\xt -1)(\xt-2) \;  \omega^R_{j,k} \\
        \frac{1}{2}(\xt+1)\; \xt\; \omega^C_{j,k} -\xt\; (\xt-2)\; \omega^R_{j,k} ,&\hspace*{-0.5cm}  \frac{1}{2}\leq \xt < \frac{3}{2}\\
         \frac{1}{2}\xt \; (\xt-1) \; \omega^R_{j,k} &\hspace*{-0.2cm}  \frac{3}{2}\leq \xt < \frac{5}{2}\\
         0 &  \hspace*{-0.2cm} \xt > \frac{5}{2}  \\
         \end{array}\right. \\
  \]

\noindent{B. 5th-order Hermite interpolation}\\
If one where to use standard Lagrange Polynomial interpolation when
mapping metric data from the grid to the particle positions, the particle
would see a continuous but non-differentiable metric when crossing grid
lines. To avoid the extra noise caused by this, we have implemented a
5th order Hermite interpolation scheme (following \cite{timmes00a}) for the
mapping of metric quantities from the grid to the particle positions.

Even in the presence of hydrodynamical shocks, the metric will be at least
twice differentiable (i.e.\ $C^2$). By using Hermite interpolation we ensure
that the interpolated values are $C^2$ across grid boundaries. In one dimension,
on the interval $[x_i,x_{i+1}]$, we therefore want to define an interpolating
function, $f(x)$, that has the following properties:
\begin{eqnarray}
f(x_i) = f_i = C_1, & \mbox{\vspace{2em}} & f(x_{i+1}) = f_{i+1} = C_2, \nonumber \\
f'(x_i) = f'_i = C_3  & \mbox{\vspace{2em}} & f'(x_{i+1}) = f'_{i+1} = C_4, \\
f''(x_i) = f''_i = C_5  & \mbox{\vspace{2em}} & f''(x_{i+1}) = f''_{i+1} = C_6. \nonumber
\end{eqnarray}
As we have six conditions to impose, $f(x)$ needs to be at least a 5th order
polynomial. Introducing
\begin{equation}
\Delta x = x_{i+1}-x_i
\end{equation}
and
\begin{equation}
\tilde{x} = \frac{x-x_i}{\Delta x}
\end{equation}
we can write the interpolating quintic Hermite polynomial as
\begin{eqnarray}
H_5(\tilde{x}) & = & f_i\,\psi_0(\tilde{x})+f_{i+1}\,\psi_0(1-\tilde{x}) \nonumber \\
       & & +f'_i\Delta x\,\psi_1(\tilde{x})+f'_{i+1}\Delta x\,\psi_1(1-\tilde{x}) \\
       & & +f''_i\Delta x^2\,\psi_2(\tilde{x})+f''_{i+1}\Delta x^2\,\psi_2(1-\tilde{x}),
             \nonumber 
\end{eqnarray}
where the conditions on the function values and derivatives determine the
3 quintic Hermite basis functions
\begin{eqnarray}
\psi_0(\tilde{x}) & = & -6 \tilde{x}^5 +15 \tilde{x}^4 - 10 \tilde{x}^3 + 1, \nonumber \\
\psi_1(\tilde{x}) & = & -3 \tilde{x}^5 +8 \tilde{x}^4 - 6 \tilde{x}^3 + \tilde{x}, \\
\psi_2(\tilde{x}) & = & \frac{1}{2} ( -\tilde{x}^5 +3 \tilde{x}^4 - 3 \tilde{x}^3 + \tilde{x}^2 ). \nonumber
\end{eqnarray}
As we do not know the values of the first and second derivatives of the metric
quantities at $x_i$ and and $x_{i+1}$, we approximate these by fourth order
finite differences as
\begin{eqnarray}
f'_i & = & \frac{f_{i-2}-8 f_{i-1}+8 f_{i+1}-f_{i+2)}}
                   {12 \Delta x}, \\
f''_i & = & \frac{-f_{i-2}+16 f_{i-1}-30 f_i+16 f_{i+1}-f_{i+2}}
                    {12 \Delta x^2},
\end{eqnarray}
and similarly for the point $x_{i+1}$ with the stencil shifted by one.
In one dimension the stencil for Hermite 5 interpolation thus becomes a
six point stencil from $x_{i-2}$ to $x_{i+3}$ where the point, $x$, to be
interpolated to lies in the interval $[x_i, x_{i+1}]$.

In three dimensions the Hermite interpolation stencil consists of the 216
points in the 6x6x6 cube defined by the corners $(x_{i-2},y_{i-2},z_{i-2})$ and
($x_{i+3},y_{i+3},z_{i+3})$. The interpolation to point $(x,y,z)$ then
proceeds in principle as 36 one dimensional interpolation in the $z$-direction
to the points in the square defined by $(x_{i-2},y_{i-2},z)$ to
$(x_{i+3},y_{i+3},z)$, then another six interpolations in the $y$-direction
to the points on the line from $(x_{i-2},y,z)$ to $(x_{i+3},y,z)$ and
finally a last interpolation in the $x$-direction to the point $(x,y,z)$.

In practice, however, we have prederived expressions for the weights of
all 216 points in the three dimensional stencil, so when we know which
point we have to interpolate to, we calculate the weights and
then do the interpolations in all three directions in one go. This has
the advantage, that we can reuse the weights for each function we have
to interpolate to the same point.

\subsection{Initial conditions and Artificial Pressure Method (APM)}
\label{sec:APM}
\begin{figure*}
   \centering
   \hspace*{0cm}\includegraphics[width=1.\columnwidth]{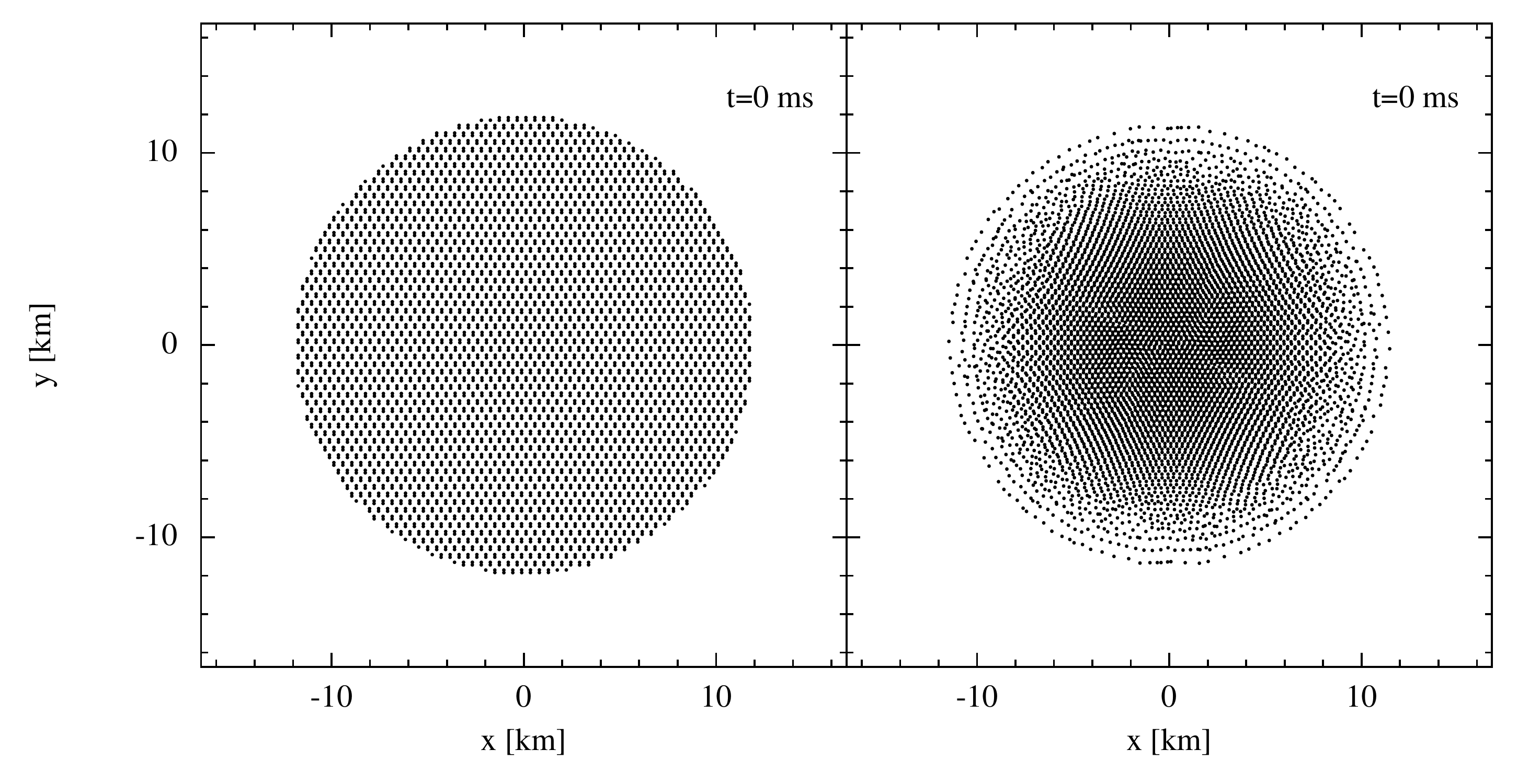} 
   \caption{Particle distribution ($|z|<0.5$) according to a uniform hexagonal lattice (left) and according to
   the Artificial Pressure Method (APM), see main text for more details.}
\label{fig:NS_APM}
\end{figure*}

Apart from the shock test described in Sec.~\ref{sec:sod}, all other tests in these papers
are concerned with the evolution of neutron stars. The initial neutron star profiles are obtained by
solving the Tolman-Oppenheimer-Volkoff (TOV) equations \cite{tolman39,oppenheimer39}. In 
setting up our initial configurations we have to take into account a peculiarity of SPH: its sensitivity
to particles of different masses (Newtonian) or baryon numbers (relativistic case). Ideally, one would
like to have initial particle distributions that a) are very regular (for a more quantitative definition of
this property see Sec. 2 in \cite{rosswog15c}), b) do not contain preferred directions (which simple
lattices usually do) and c) have equal masses/baryon numbers, i.e. the information about the density
structure should be encoded in the particle position distribution (rather than in the masses/baryon numbers
as is the case for regular lattices). In practice it can become a non-trivial task to set up particle distributions
that fulfil these properties. It should be noted, however, that in particular stiff EOSs (e.g. $\Gamma=2.75$) 
with their nearly uniform densities can still be handled with a uniform lattice. For the resolutions
shown in this paper, a uniform setup results in baryon number ratios of $\sim 8$ between center 
and the resolvable neutron star surface; which is perfectly acceptable. 
For $\Gamma=2.0$, however, this ratio becomes much larger ($>10^4$) and here a more sophisticated 
setup is beneficial.\\
For such a setup,  we modify the ``Artificial Pressure Method'' (APM) that has recently been suggested in the context
of the  Newtonian SPH code MAGMA2 \cite{rosswog20a} for the case of relativistic TOV-stars.
The main idea of the APM method is to distribute equal mass/baryon number particles, measure their
current density according to Eq.(\ref{eq:N_sum}) and then define an ``artificial pressure'' based on
the relative deviation  between the measured density  and the desired profile density. This artificial pressure
is used in a momentum-type equation similar to Eq.(\ref{eq:dSdt_hydro}) to drive the equal mass particles iteratively 
into positions where they minimize the deviation from the desired density profile. What we use here is a
straight-forward translation of the original Newtonian method. Here we briefly summarize the method and
refer to the original paper for more details and tests.\\
Specifically, we follow the following steps:
\bi
\i Distribute the initial guess of the particle positions. To this end we have implemented  a regular cubic 
and a hexagonal lattice. The particles are placed in a sphere of radius $1.2 R_{\rm NS}$, where
$R_{\rm NS}$ is the radius of the TOV solution. The particles outside $R_{\rm NS}$ serve as boundary 
particles in the iteration process and are discarded once the iteration process has converged.
\i In the next step we assign the artificial pressure $\Pi_a$ to particle $a$ according to
\be
\Pi_a \equiv {\rm max} \left[ 1 + \frac{N_a - N^{\rm TOV}(\vec{r}_a)}{N^{\rm TOV}(\vec{r}_a)}, 0.1\right]
\ee
and use it for the 
\i position update $\vec{r}_a \rightarrow \vec{r}_a + \delta \vec{r}_a^{\rm APM}$ where
\be
\delta \vec{r}_a^{\rm APM} = -\frac{1}{2} h_a^2 \nu \sum_b \frac{\Pi_a + \Pi_b}{N_b} \nabla_a W_{ab}(h_a).
\ee
As outlined above, this is a straight-forward translation of the Newtonian method, the details of which can be
found in Sec. 3.1 of \cite{rosswog20a}. This update procedure tries to minimize the density error for the given
baryon mass $\nu$ of all the particles, but it does not consider the regularity of the particle distribution. To achieve
a good compromise between good density estimate (for the same $\nu$) and a locally regular particle distribution
we add a regularization term similar to \cite{gaburov11}
\be
\delta \vec{r}_a^{\rm reg} =  h_a^4 W_{ab}(h_a) \hat{e}_{ab},
\ee
so that the final position correction is
\be
\delta \vec{r}_a= (1 - \zeta) \delta \vec{r}_a^{\rm APM}  + \zeta \delta \vec{r}_a^{\rm reg}.
\ee
After some experimenting we settled for a value of $\zeta= 0.1$ for the regularization contribution.
\i The SPH form of the hydrodynamic equations, Eq.~(\ref{eq:dSdt_hydro}), has excellent momentum
conservation properties, but the gravitational acceleration terms that are calculated on a mesh
and interpolated back to the particle positions can introduce a small momentum violation if the
particles are not perfectly symmetrically distributed. As a thought experiment think of the star being
composed of only two SPH particles: even if the accelerations are exactly the TOV values, this
will result in a non-zero total momentum change unless the particles are symmetric with respect
to the centre of the star. Therefore, we enforce perfect symmetry
after each position update, simply by assigning to each of the first half of the particles a ``mirror
particle'' that is symmetric with respect to the origin.
\i To monitor the convergence, we measure the average density error and once it has not improved
for 20 trial iterations, we consider the particle distribution as converged. 
\i Once this stage has been reached, we improve the agreement with the TOV-solution by {\em now}
adjusting the particle masses in an iterative process. This leads to final ratios in the SPH particle
baryon numbers of a few, which is perfectly acceptable. The exact ratio depends on how centrally
condensed the stellar model is (i.e. on the equation of state) and we find ratios of $\sim2$ for a stiff,
$\Gamma= 2.75$, and  ratios of $\sim6$ for a softer, $\Gamma= 2.0$, equation of state 
(compared to ratios $>10^4$ for straight-forward lattice setup).
\ei 
Note that with our setup we try to closely approximate the density distribution, the exact baryon mass
is not actively enforced and can therefore be used as a consistency check. We find that it agrees
very well with the one from the TOV profile, typically to $\sim 0.2$ \% for stars with a few hundred thousand 
particles.
An example of initial particle distributions ($|z| < 0.5$, 100k particles) of a $\Gamma= 2.0$ equation of state is shown
in Fig.~\ref{fig:NS_APM}. The left panel shows a hexagonal lattice  while the right panel shows a setup
according to the APM (max./min. baryon number for this case $\sim 6$). 

\subsection{Code Implementation}
\label{sec:implementation}
Apart from the used \Mcl thorn~\cite{Brown:2008sb} (described in the spacetime section above)
our code has been written entirely from scratch in modern Fortran (with elements up to Fortran 2008).
Since \SpB has been written alongside the high-precision SPH code \Ma \cite{rosswog20a}, both codes share some 
modules such as the kernel calculation and parts of the tree-infrastructure for the neighbour search. \SpB will be
developed  further in the near future. In its current stage it is OpenMp parallelized and it takes about
10 hours of wall clock time on an Intel Cascade Lake Platinum 9242 (CLX-AP) node to evolve 1 million 
particles together with a $300^3$ uniform mesh  for a physical time of 1 ms. For now, a uniform mesh 
is implemented, but this may be improved in the future.

\section{Tests}
\label{sec:tests}
All tests shown here are performed with the full 3+1 dimensional hydrodynamics code. We have
run a very large set of experiments where we evolved TOV neutron stars (hydrodynamics and spacetime) 
and we monitored how close the solution remained to the 1D TOV-solution for different combinations of our
numerical choices. After these tests we settled on the following default choices:  the sequence 
$MLS-MLS-MLS-M_3$ (from $fs=3$ to 0) for the P2M-mapping and  5th order Hermite interpolation for the
P2M-mapping.
We use $\alpha_0= 0.1$ as minimum and 1.5 as the maximum dissipation value. But note that a number of other combinations yield
very similar results. For example, $M^{'''}_6-M^{'''}_6-M_4-M_3$ works nearly as well and is computationally
slightly cheaper, though the P2M-step is only a moderate fraction of the overall computational time.\\
Our code uses units with $G=c=1$ and masses are measured in solar units.
Unless units are explicitly
provided, all parameters given for initial data are in code units.
These are often useful because the initial conditions of many tests that we
show are given in the literature also in these units. However, for the physical
results related to neutron stars, we prefer to use physical (cgs-)units, 
but we believe that this use of units should not lead to any confusion.

\subsection{Relativistic shock tube}
\label{sec:sod}
\begin{figure}
   \centering
   \includegraphics[width=1.\columnwidth]{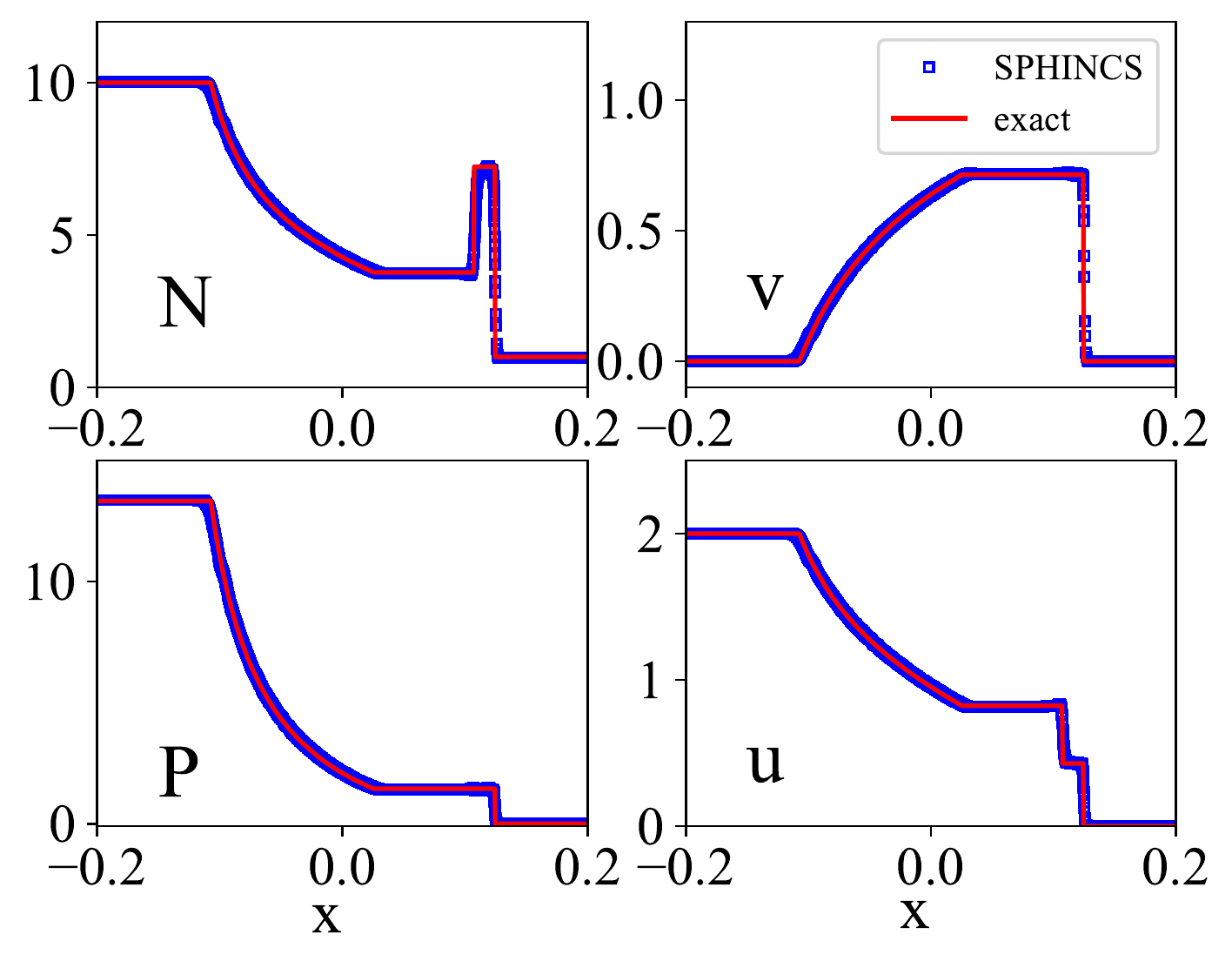} 
   \caption{Result of a 3D relativistic shock tube (initial particle spacing on the left $\Delta x_L=0.0005$) at t= 0.15, 
   numerical results are shown as blue squares, the exact solution is shown in red.}
\label{fig:rel_Sod}
\end{figure}

\begin{figure}
   \centering
   \includegraphics[width=1.\columnwidth]{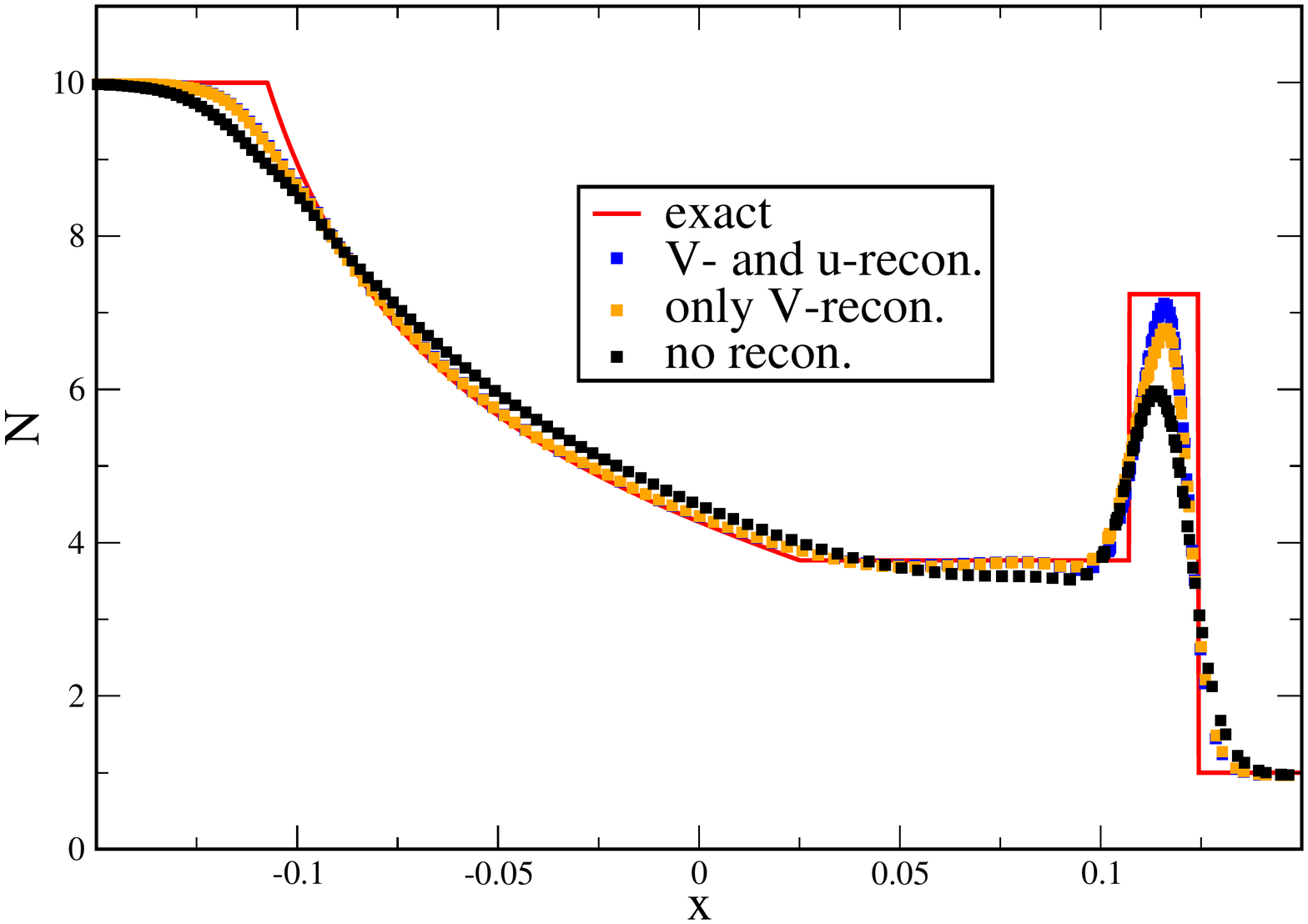} 
   \caption{Result of a low resolution 3D relativistic shock tube (initial particle spacing on the left $\Delta x_L=0.003$) at t= 0.15.
   The exact solution is shown in red, the numerical solution without any reconstruction in black,
   with reconstruction in only $V$ in orange and with reconstruction in both $V$ and $u$ in blue.}
\label{fig:Sod_recon}
\end{figure}

In this first test we scrutinize the ability of our full-GR code to correctly
reproduce the special-relativistic hydrodynamics limit.
The test is a relativistic version of ``Sod's shocktube'' \cite{sod78} which  
has become a widespread benchmark for relativistic hydrodynamics codes 
\cite{marti96,chow97,siegler00a,delzanna02,marti03}. The test uses a polytropic 
exponent $\Gamma=5/3$ and as initial conditions 
\be
\left[ N, P \right]=   \left\{
    \begin{array}{ll}
         \left[10,\frac{40}{3}\right], & {\rm for \;}  x<0\\
   	 \left[1,10^{-6}\right]          & {\rm for \;} x \ge 0,
   \end{array}
  \right.
\ee
with velocities initially being zero everywhere.
We place  particles with equal baryon numbers on close-packed lattices as described in \cite{rosswog15b},
so that on the left side the particle spacing  is $\Delta x_L= 0.0005$ and we have 12 particles in both y- 
and z-direction. This test is performed with the full 3+1 dimensional code, but  using a fixed Minkowski metric.
The result at $t= 0.15$ is shown in Fig.~\ref{fig:rel_Sod} with the \SpB results
marked with blue squares and the exact solution \cite{marti03} with the red line. Overall there is very good 
agreement  with practically no spurious oscillations. 
To illustrate the effect of the reconstruction in the artificial dissipation we repeat this 3D test
at low resolution ($\Delta x_L= 0.003$), once without, once with reconstruction in only $V$
and once with reconstruction in $V$ and $u$, see Fig.~\ref{fig:Sod_recon}.

\subsection{Hydrodynamic evolution of neutron star in a static metric (``Cowling approximation'')}
\label{sec:cowling_osc}
\begin{figure}
   \centering
   \includegraphics[width=1.\columnwidth]{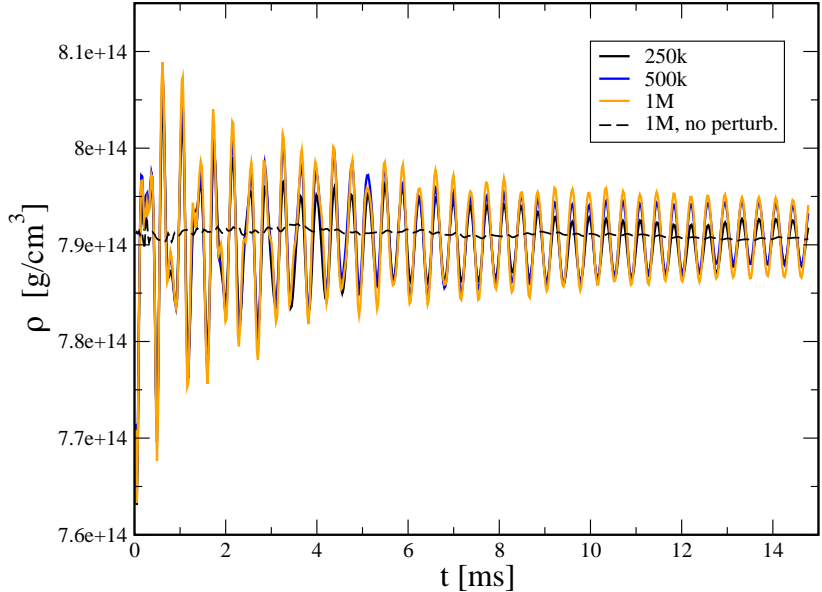} 
   \caption{Central density oscillations of three stars (250k, 500k and 1M SPH particles)
                 in Cowling approximation with a $\Gamma=2.0$ polytropic equation of state.
		 The oscillations have been triggered by a small velocity perturbation 
		 $\delta v_0= 0.005$. We also show a case where no explicit perturbation 
		 was applied (black dashed line).}
\label{fig:Cowling_oscillations}
\end{figure}

\begin{figure}
   \centering
   \includegraphics[width=1.\columnwidth]{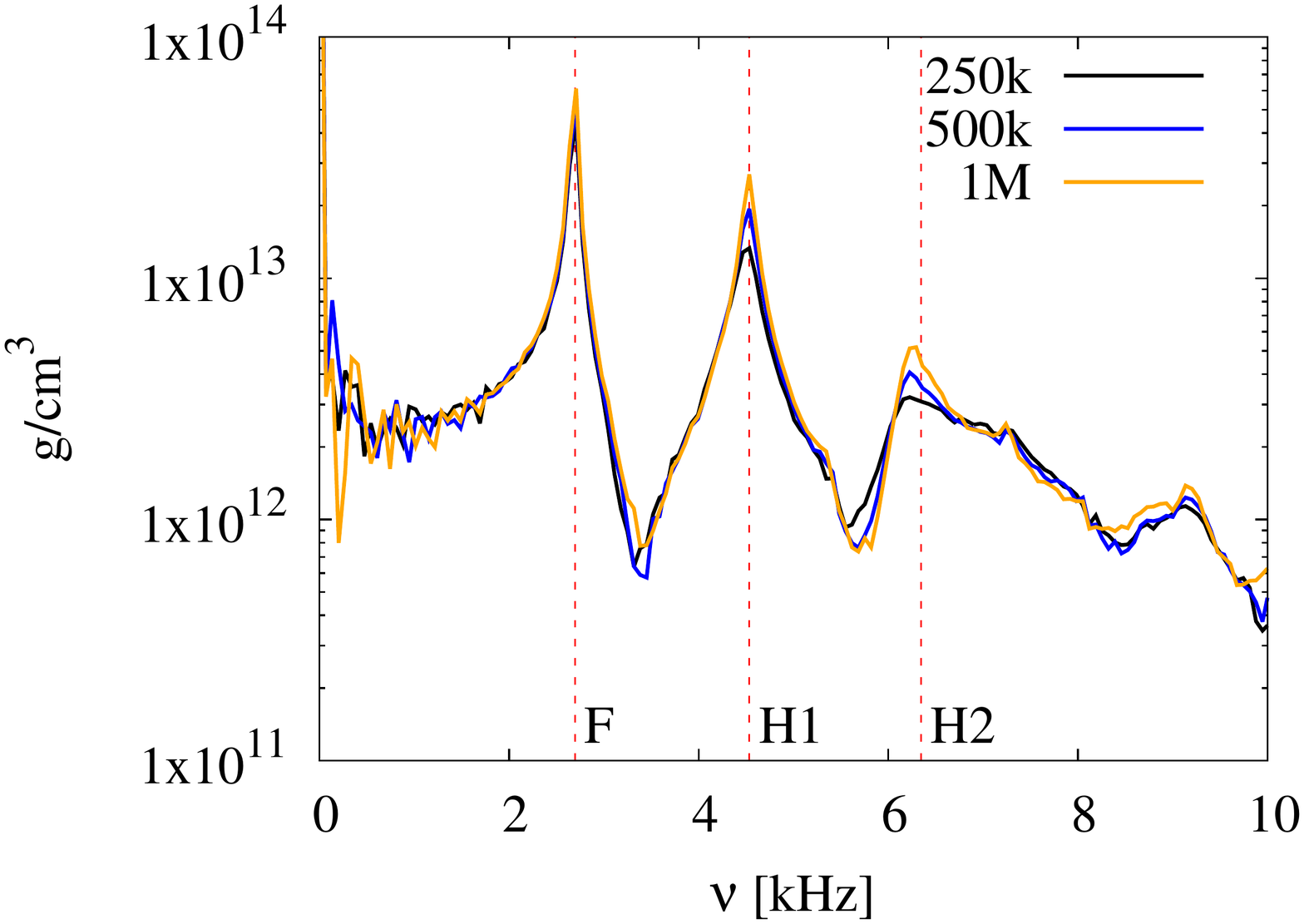}
   \caption{Fourier spectrum of the central density oscillations of the three
            stars shown Fig.~\ref{fig:Cowling_oscillations} that have been evolved
            in the "Cowling approximation" (matter is evolved, but spacetime held fixed).
             Also indicated with the red dashed vertical lines are the fundamental normal
            mode frequency (F) and the next two higher mode frequencies (H1,
            H2) as determined by  \protect\cite{font02} in a 3D study.}
\label{fig:Cowling_spectrum}
\end{figure}

After testing the {\em special}-relativistic performance of the hydrodynamic terms 
in the previous shock test, we next test the {\em general}-relativistic hydrodynamics 
by evolving the matter variables of a neutron star  while keeping the metric fixed (``Cowling
approximation''). The purpose of this test is two-fold: a) it should demonstrate that the 3D
star remains close to the initial solution that has been found by solving the 1D TOV-equations
and b) we will measure oscillation frequencies and compare them to results from the literature.
To enable a straight-forward  comparison we follow here the setup of \cite{font02} who also provide their 
results for the oscillation frequencies.
We model a 1.40 \Msun (gravitational mass) neutron star by solving the TOV equations 
with a $\Gamma=2.00$ polytropic exponent, a prefactor of $K= 100$ in the polytropic
equation of state, $P= K n^\Gamma$, and a central density of $\rho_c= 1.28 \times 10^{-3}$.\\
We set up initial TOV stars according to the APM described in Sec.~\ref{sec:APM} at three 
different resolutions: 250k, 500k and 1M particles. Note that in Newtonian SPH one
usually ``relaxes'' a star to find its true numerical equilibrium. This is usually done, see e.g.
\cite{rosswog99}, by setting up the particles as closely to the hydrostatic equilibrium as
possible and then let them evolve with some extra-dissipation, so that they can settle 
locally into an ideal particle configuration. We do not perform such a relaxation step here, but
start directly with the stars from the APM setup. Therefore, in the initial phase
the particles will try to further optimize their local arrangement in addition to a possible
bulk motion.  To set the star into oscillation we apply a small radial perturbation 
\be
\delta v^r=  \delta v_0  \sin\left(\frac{\pi r}{R}\right),
\label{eq:d_vr}
\ee
where $\delta v_0= 0.005$. \\
The evolution of the central densities of these stars over $\approx 15$ ms is shown in Fig.~\ref{fig:Cowling_oscillations}.
Overall, the stars at all resolutions stay close to the initial TOV solution and stably oscillate around it without noticeable
systematic drift. The oscillations are somewhat damped by numerical viscosity, but notably less so
with increasing resolution.\\
We also measure the oscillation frequencies and present the resulting Fourier
spectrum in Fig.~\ref{fig:Cowling_spectrum}. We find excellent agreement with
the values for the fundamental normal mode (F: 2.696 kHz) and the first two
overtones (H1: 4.534 kHz, H2: 6.346 kHz) determined in \cite{font02} using a
3D Eulerian high resolution shock capturing code. The spectrum agrees well
among the three resolutions and as expected, the peaks get sharper and have
higher amplitudes at higher resolution. Higher order overtones are excited at
a too low amplitude to be visible in the spectrum.

\subsection{Stable neutron star with dynamical spacetime evolution}
\begin{figure}
   \centering
   \includegraphics[width=1.\columnwidth]{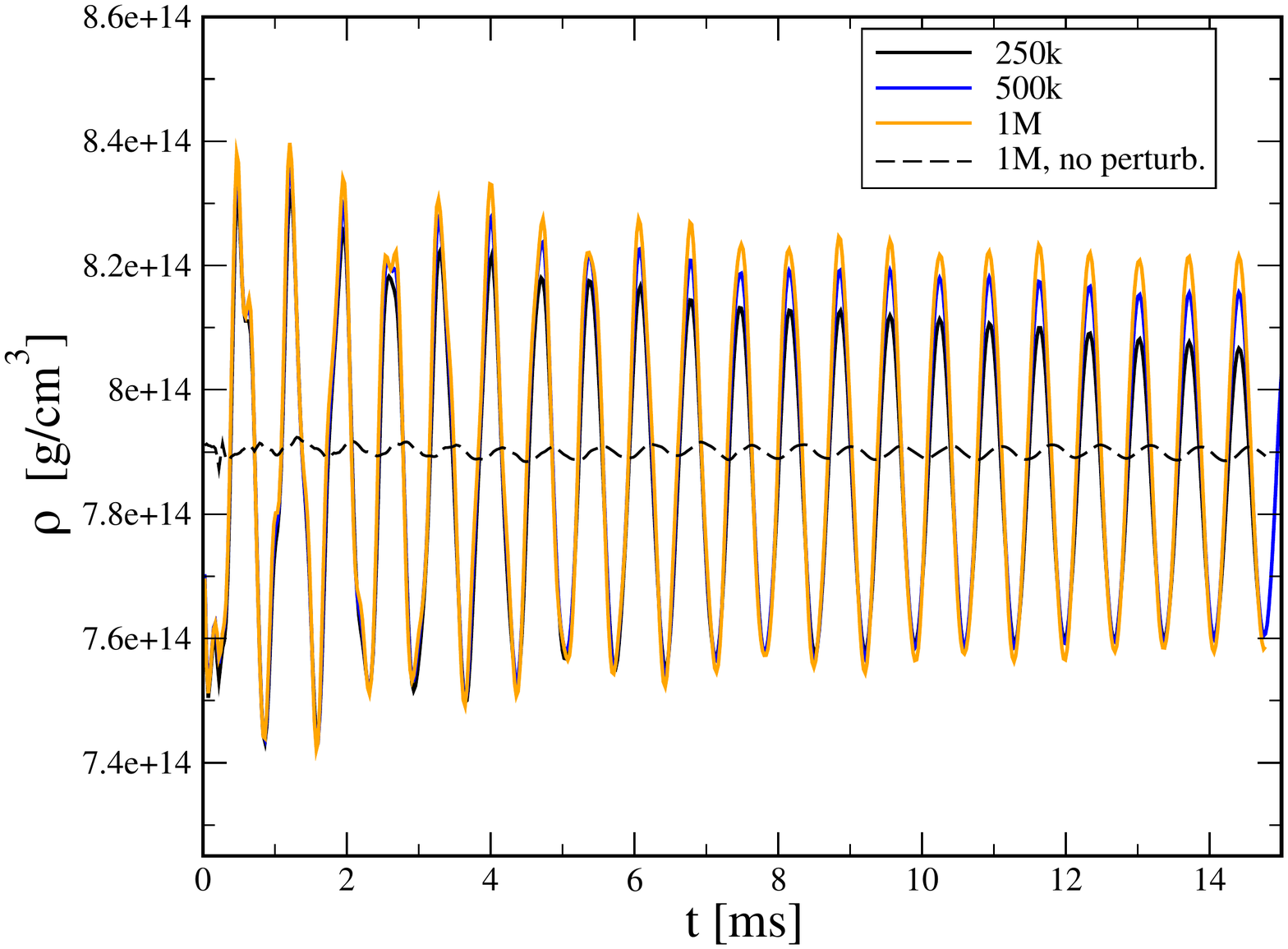} 
   \caption{Central density oscillations in the full matter+spacetime evolution  of a neutron star.
                 The initial model is  set up according to the TOV equations with a polytropic exponent $\Gamma= 2.00$
                 The used SPH particle numbers are indicated in the 
		 legend (``k'': thousands, ``M'': million),  the spacetime was evolved on a grid 
                 covering a volume of $[-30,30]^3$ (in code units; 1 length code unit= 1.47676 km).
                 See main text for more details. The solid lines show cases where the oscillations 
                 were triggered with a small velocity perturbation, the thin dashed line shows the result for 1 million particles 
                 where no explicit velocity perturbation was applied and the oscillations are triggered exclusively by truncation
		 error.}
\label{fig:rho_fullEvol}
\end{figure}

\begin{figure}
   \centering
   \includegraphics[width=1.\columnwidth]{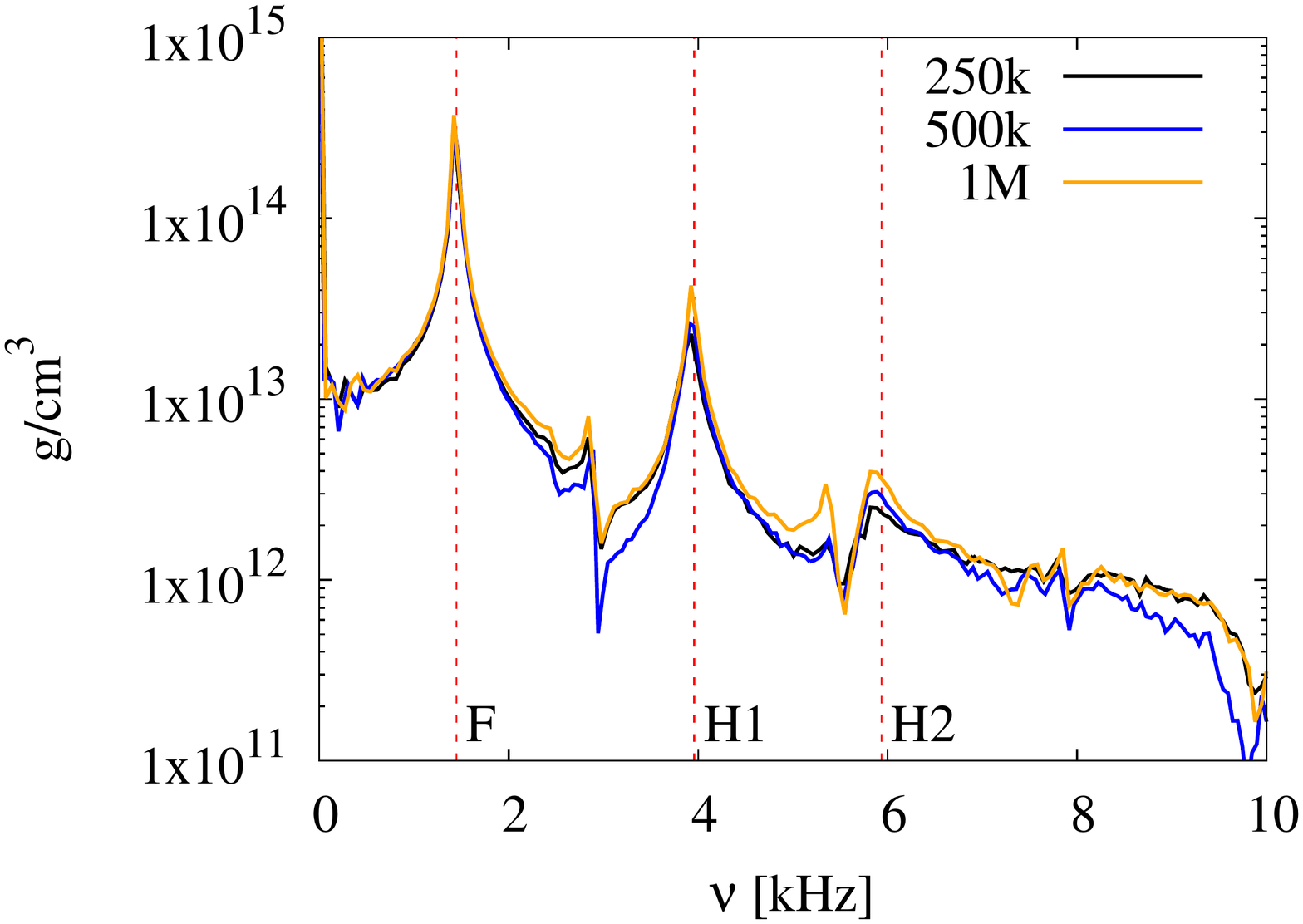}
   \caption{Fourier spectrum of the central density oscillations of neutron stars 
                 that were self-consistently evolved together with the spacetime, see
                 Fig.~\ref{fig:rho_fullEvol}. Also indicated
                 with the red dashed vertical lines are the fundamental normal
                 mode frequency (F) and the next two higher mode frequencies (H1,
                 H2) as determined by  \protect\cite{font02}.}
\label{fig:Full_spectrum}
\end{figure}

As the next step, we take the configuration from the previous test, but now also evolve
the spacetime dynamically, i.e. we are testing the general relativistic hydrodynamics, the spacetime
evolution and their coupling. During the subsequent numerical evolution, the neutron star should
remain stable and close to the initial TOV setup. As a further test, we measure again the oscillation
frequencies of the star and compare them against the results published in \cite{font02}.\\
We use a setup very similar to the previous test and use in particular (unrelaxed) initial 
configurations with 250k (grid resolution $126^3$), 500k ($151^3$) and 1M particles ($191^3$ grid points).
The number of grid points has been chosen so that the average number of particles per grid cell
is approximately the same ($\approx15$).  We slightly perturb the stars 
according to Eq.~(\ref{eq:d_vr}). The evolution of the central densities are shown in Fig.~\ref{fig:rho_fullEvol}.
Again the stars oscillate stably around the initial TOV central density and with only moderate decrease
in oscillation amplitude due to dissipative effects. As expected, and as seen before, the dissipation decreases
further with increasing resolution.\\
We measure again the oscillation frequencies and present them in Fig.~\ref{fig:Full_spectrum}. Once more, 
we find excellent agreement with the values for the
fundamental normal mode (F: 1.450 kHz) and the first two overtones 
(H1: 3.958 kHz, H2: 5.935 kHz) from \cite{font02}. As in the Cowling case,
higher overtones are excited at too low amplitudes to be seen reliably in the
spectrum. However, there might be a small hint of H3 at 7.812 kHz.

\begin{figure*}
   \centering
   \includegraphics[width=\columnwidth]{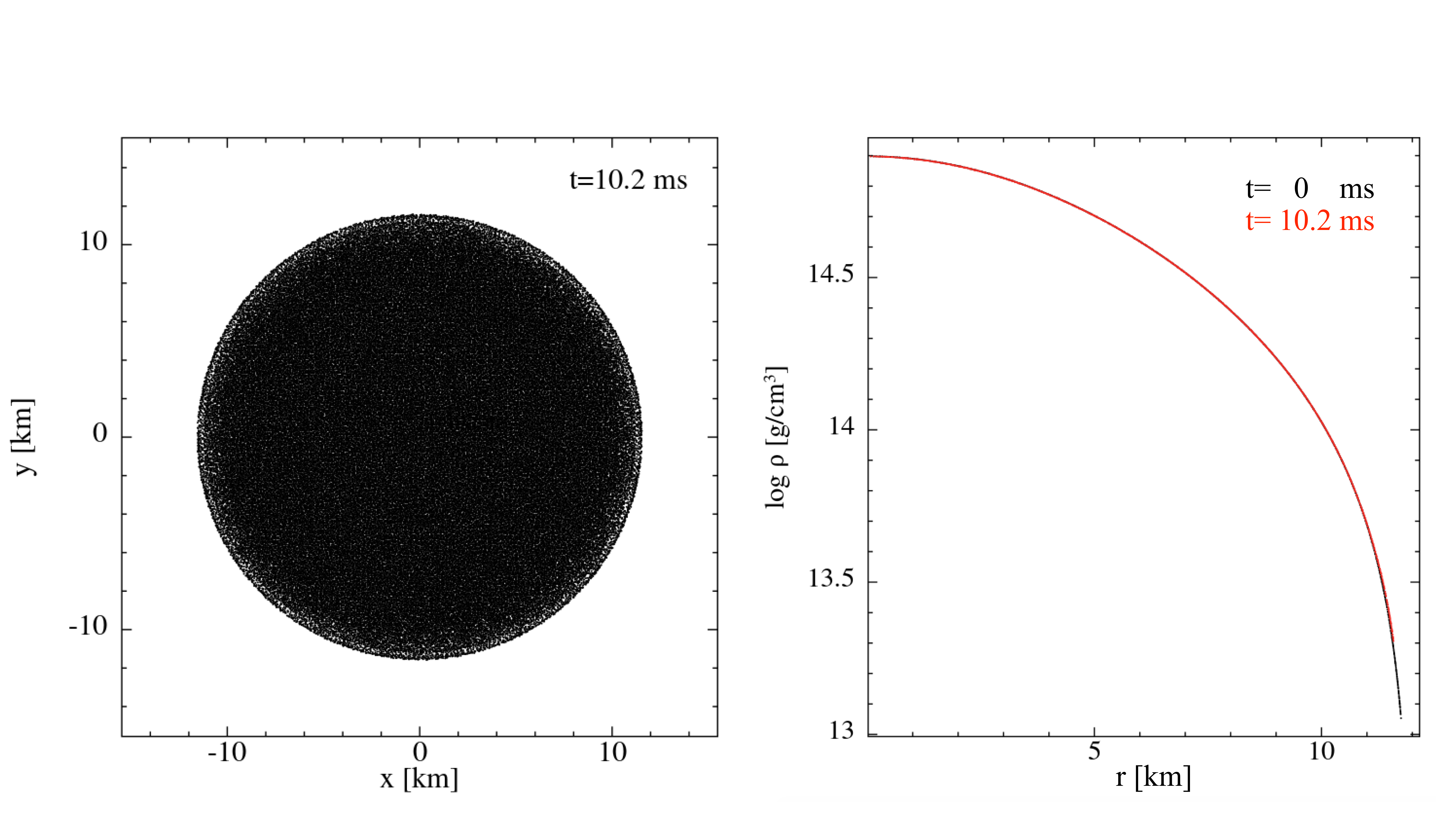}
   \caption{TOV neutron star ($\Gamma=2.0$) after it has been evolved (hydrodynamics and spacetime) for 10.2 ms. Only
   in the surface layers have the particle positions slightly adjusted, otherwise the star remained essentially perfectly
   on its initial condition.  Note in particular that the stellar surface --which does not need any special treatment in our approach-- 
   has remained perfectly well-behaved. }
\label{fig:TOV_2.0_0_10ms}
\end{figure*}

In Fig.~\ref{fig:TOV_2.0_0_10ms} (left panel) we show the particle distribution of a $\Gamma=2.0$ star (no initial velocity perturbation;
shown as black dashed line in Fig.~\ref{fig:rho_fullEvol})
after it has been evolved (hydrodynamics and spacetime) for $10.2$ ms. Note that, contrary to Eulerian General Relativity approaches, the 
neutron star surface does not pose any particular challenge for our numerical method: the surface remains sharp and perfectly
well-behaved. In Fig.~\ref{fig:TOV_2.0_0_10ms} (right panel) we show the radial
structure of the density at time $0$ ms and $10.2$ ms. Note that during the 10
ms evolution the particles at the surface have slightly adjusted their 
positions compared to our initial setup and sit now at a slightly lower radius, but apart from that the 
radial density structure of the star after 10 ms is practically identical to the initial condition.

\subsection{Migration of an unstable neutron star to the stable branch}
\label{sec:migration}
\begin{figure*}
   \centering
   \includegraphics[width=\columnwidth]{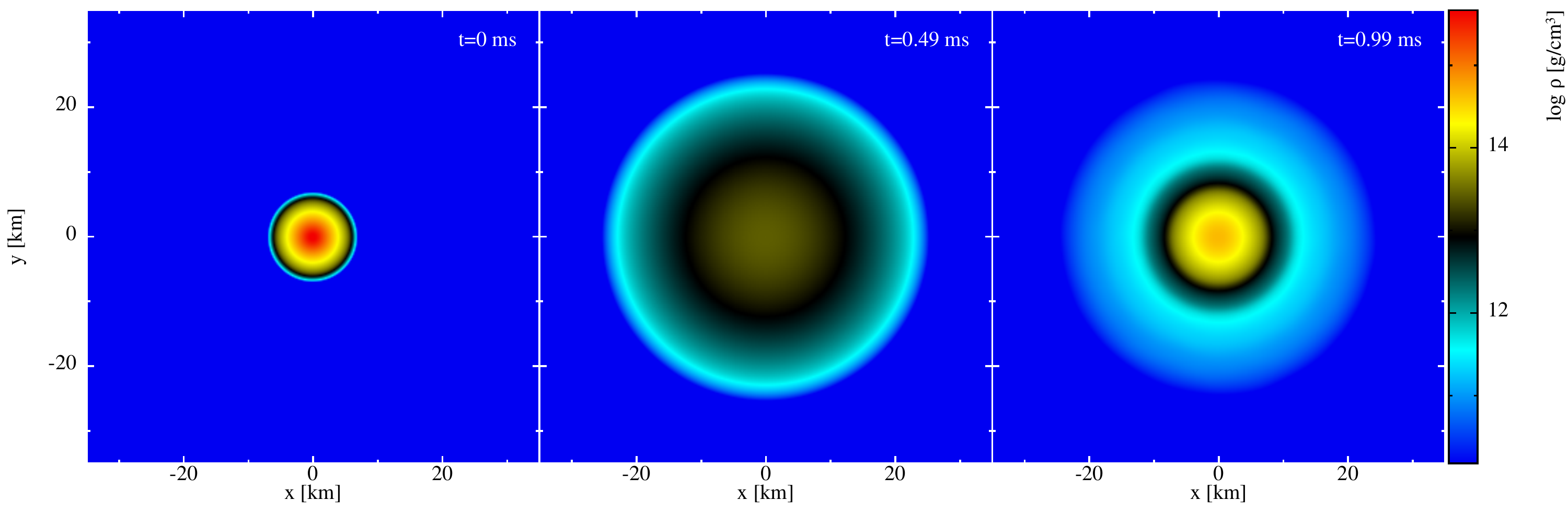}
   \caption{Density evolution in the migration test. In this test a highly relativistic neutron star is
   initially placed the unstable branch of the sequence of equilibrium stars, then undergoes
   large scale oscillations and finally settles on the stable branch. Please see main text for more details. }
\label{fig:Migration_rho}
\end{figure*}

\begin{figure}
   \centering
   \includegraphics[width=1.\columnwidth]{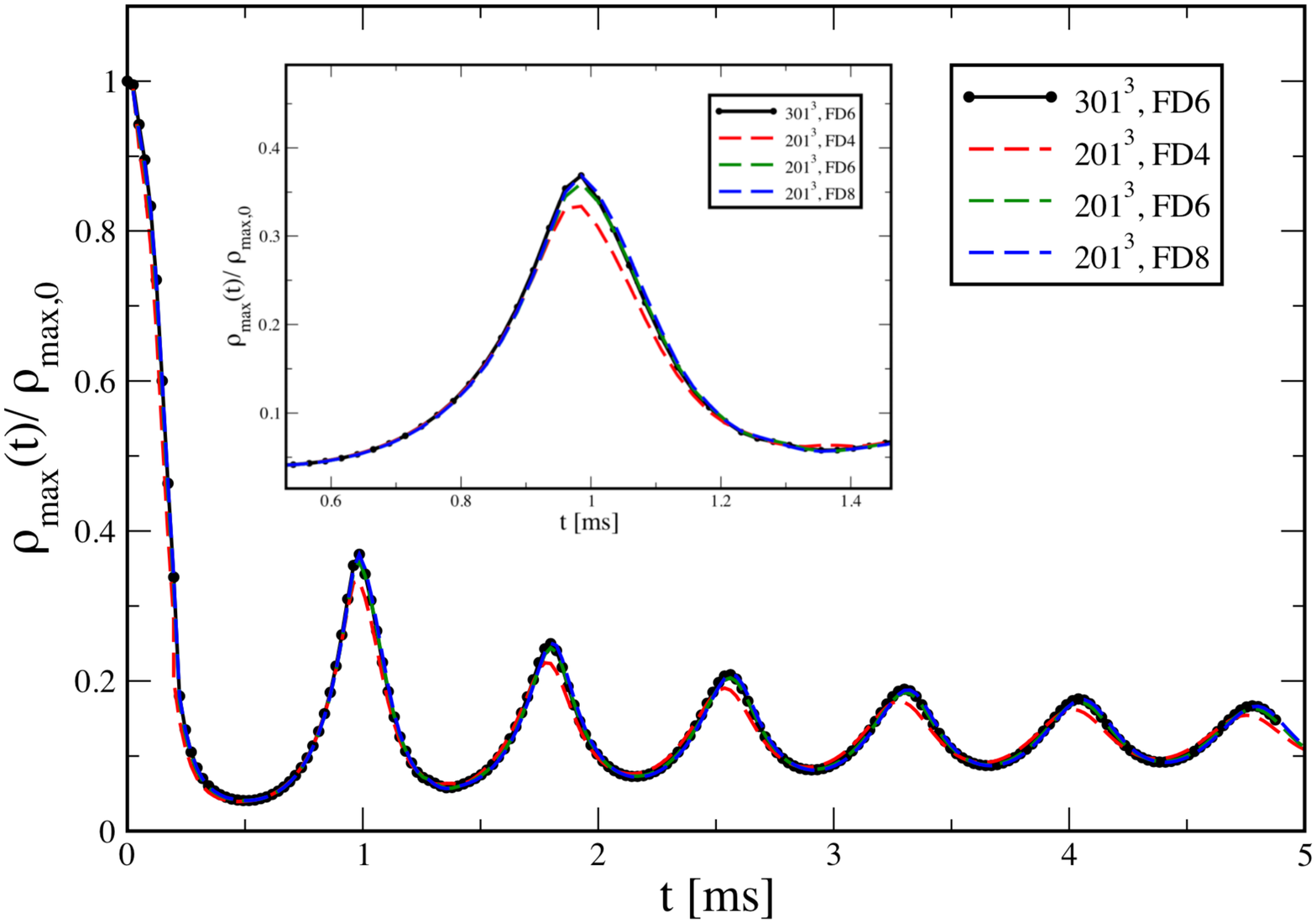}
   \caption{Evolution of the peak density (normalized to the initial peak density) in the migration test. 
   Our default in this test (black line) was performed with 1 million particles and a $301^3$ grid extending from -50 to 50 in each dimension
   and with finite differencing order 6 (``FD6''). We performed additional tests, also with 1 million particles, but with a $201^3$ grid where we varied the finite difference order
   from 4 to 8.}
\label{fig:Migration_central_rho}
\end{figure}

A more complex test case involves an unstable initial configuration
of a neutron star
 \cite{font02,cordero09,bernuzzi10}. Depending on the 
type of perturbation, such a star can either expand, collapse to a black hole or migrate
to the stable branch of the sequence of equilibrium stars. In the latter case,
the energy difference between the two configurations causes large-scale pulsations
while the star transitions to the stable branch. \\
This test is very challenging for a number of reasons. The initial neutron star  is highly relativistic 
with $\rho_c\approx 5 \times 10^{15}$ g/cm$^3$ and a central lapse
$\alpha_c < 0.3$. In the subsequent evolution the star expands by
about a factor of three in radius while its central density drops by about a factor of 30.
Thereafter it re-collapses and re-expands repeatedly with each cycle resulting
in the formation of shocks which  eject particles reaching velocities exceeding 0.6 times 
the speed of light and which unbind a non-negligible amount of the initial stellar mass. The test
is also challenging for purely numerical reasons, especially when uniform grids are
involved, since on the one hand the matter evolution should be followed far enough out
so that ejecta can be clearly separated from matter falling back and, on the other hand,
the initial star is highly centrally concentrated so that short length scales need to be 
resolved near the stellar centre.
Clearly, this complex evolution involving  strong gravity dynamically coupled 
to the hydrodynamic evolution, shock formation, matter ejection and fallback 
is far  beyond the possibilities of any linear approximation.\\
For the initial conditions we follow the setup described in \cite{bernuzzi10} and start from a solution
of the TOV-equations with a polytropic equation of state,  $P=  K n^\Gamma$ with 
exponent $\Gamma=2$ and $K=100$ and subsequently evolved using 
Eq.~(\ref{eq:poly_EOS}). With a central density of $7.993 \times 10^{-3}$ the
star has gravitational mass of 1.448 and an (isotropic coordinate) radius of $R= 5.838$.
The transition is triggered just by truncation error. We setup the star according to the 
APM with 1M particles, use a $301^3$ grid extending from -50 to 50 in each 
dimension and apply sixth order finite differencing in BSSN. The density evolution is shown in 
Fig.~\ref{fig:Migration_rho}. The star rapidly expands by about a factor of three, then recollapses, forms a shock wave that is travelling
outward and unbinding matter, recollapsing and so on. When we stop the simulation at
$t\approx 5$ ms, about 0.099 \Msun have become unbound (particles were removed at
a radius of 45). The corresponding evolution of the peak density (normalized to the initial value)
is shown in Fig.~\ref{fig:Migration_central_rho}, it agrees well with the results obtained by 
other methods \cite{font02,cordero09,bernuzzi10}. We have further performed test calculations
with only $201^3$ grid points to explore the impact of the finite difference order in the BSSN part.
The results for order four, six and eight are shown in Fig.~\ref{fig:Migration_central_rho} as red,
green and blue lines. The fourth order case shows somewhat lower peaks, which probably 
indicates that the steep central gradients are not resolved well enough. The other cases 
give nearly identical results.

\subsection{Collapse of a neutron star to a black hole}
\begin{figure}
   \includegraphics[width=1.\columnwidth]{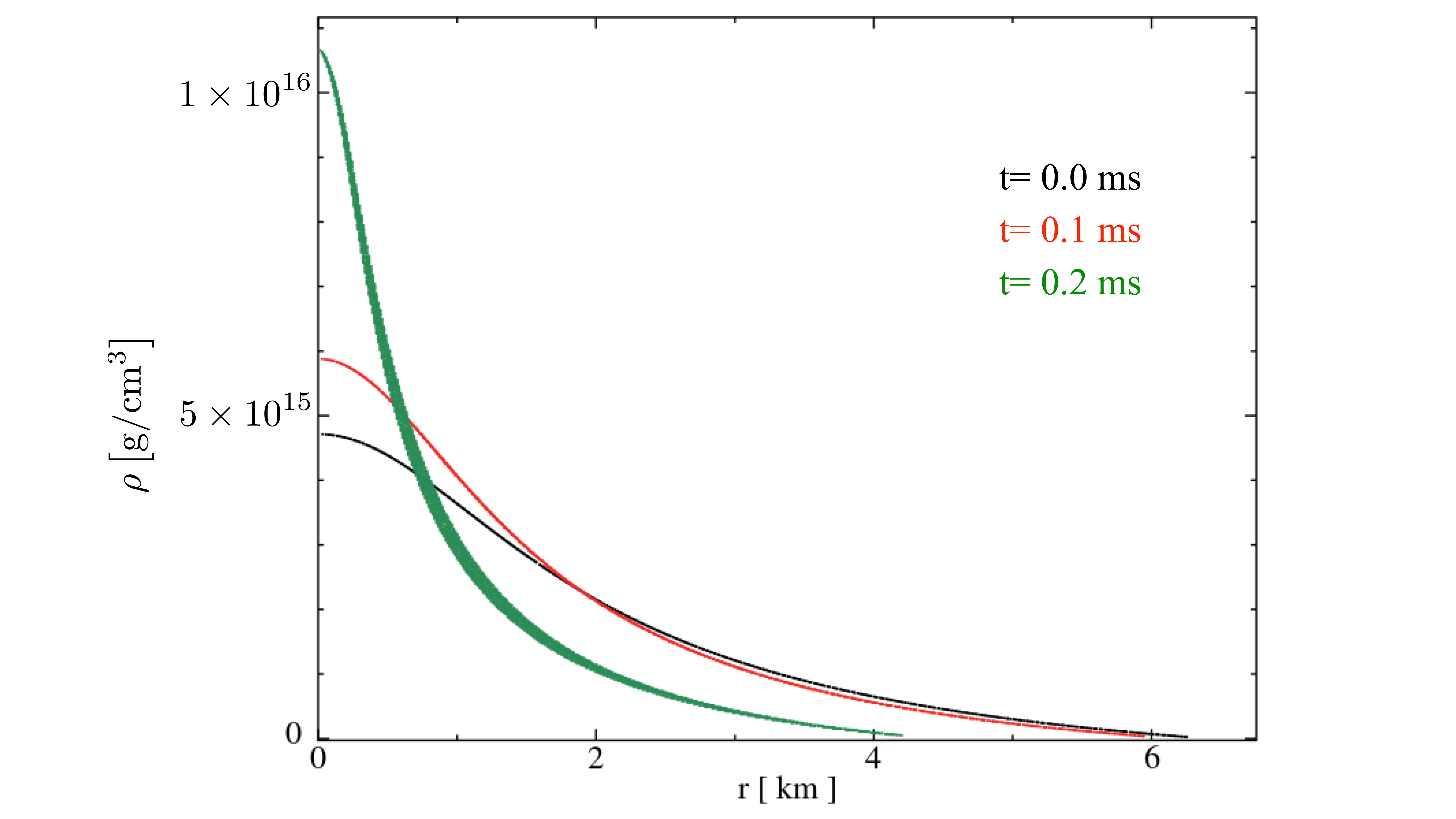}
   \caption{Density distribution in the collapsing neutron star test (initial condition is shown in black,
   the density at $t=0.1$ ms in red and at $t=0.2$ ms in green). The collapse is triggered by a small radial velocity perturbation.
   The test is performed with $401^3$ grid points and 900k SPH particles.}
\label{fig:NS_collapse}
\end{figure}

\begin{figure}
   \centering
   \includegraphics[width=1.\columnwidth]{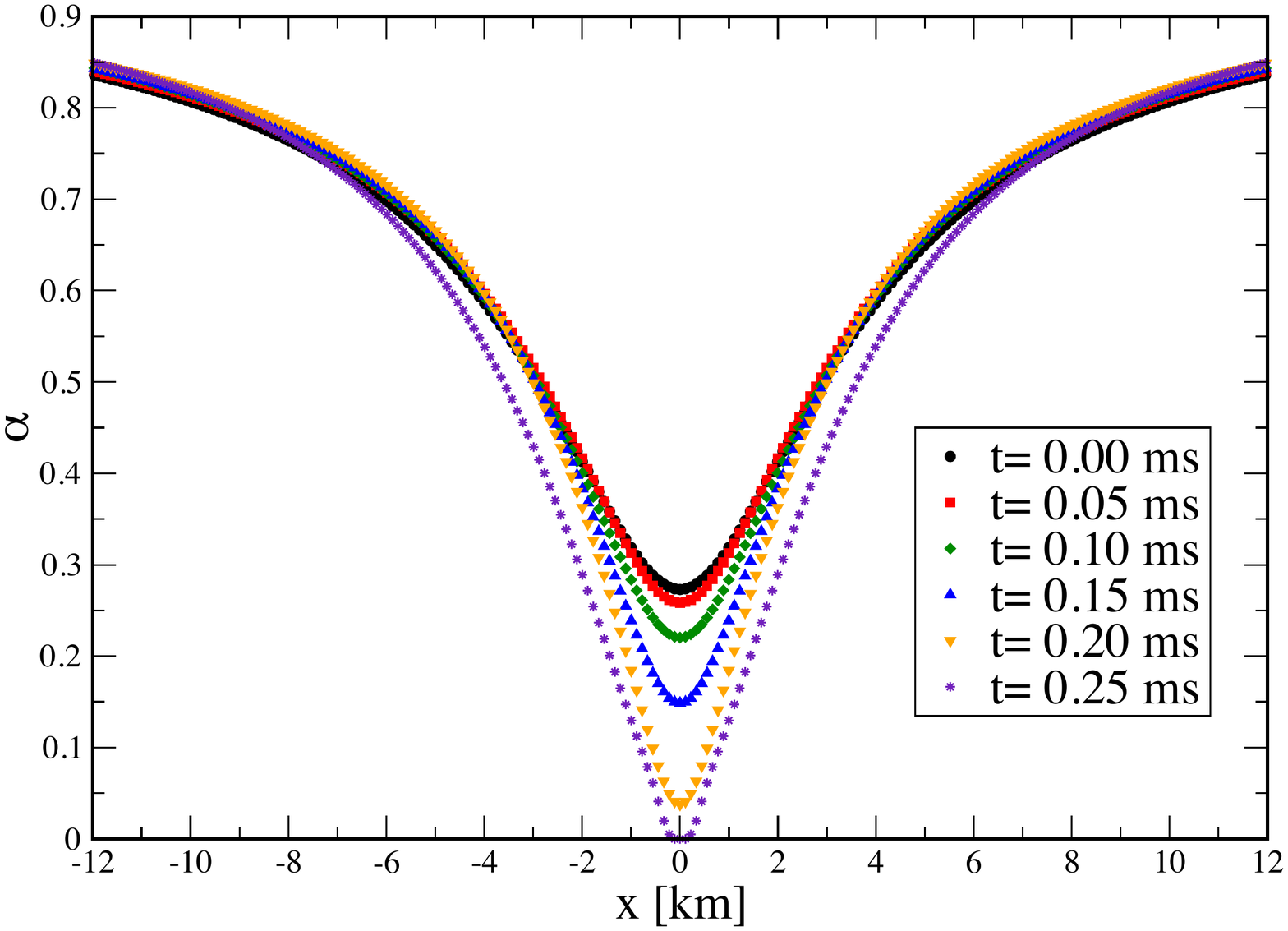}
   \caption{Evolution of the lapse function $\alpha$ along the $x$-axis for the unstable neutron star. The "collapse of the lapse"
   is characteristic for the formation of a black hole. }
\label{fig:Lapse_collapse}
\end{figure}

In this test we simulate the collapse of an unstable neutron star into a black hole.
We start from the same  initial conditions as in Sec.~\ref{sec:migration}. As mentioned there,
this configuration is unstable and depending on the perturbation, it may either -via violent
oscillations- transition to the stable branch or, otherwise, collapse into a black hole.
As demonstrated above, truncation error alone triggers the transition to the stable branch,
but only a small additional (momentum constraint violating) velocity perturbation is enough
to change the subsequent evolution and to trigger the collapse to a black hole. Similar to 
Sec.~\ref{sec:cowling_osc}, we apply a radial velocity perturbation
\be
\delta v^r= - 0.005 \sin\left(\frac{\pi r}{R}\right),
\ee
where $R$ is the stellar radius.\\
For this test, we use a $401^3$  grid with boundaries at $x_{\rm BD}=  y_{\rm BD}=  z_{\rm BD}= 15$,
6th order finite differencing and 900k SPH particles, set up according to the artificial pressure method, see 
Sec.~\ref{sec:APM}. Black hole formation goes along with a "collapse of the lapse", i.e. the lapse $\alpha$
is dropping to zero in a region inside the horizon. To prevent the hydrodynamic evolution from failing close to the forming black hole
singularity, we remove SPH particles that have a lapse value  $\alpha < \alpha_{\rm thr}= 0.03$.\\
We find that the small initial velocity perturbation triggers a rapid contraction of the neutron star which
goes along with an increase in the density, see Fig.~\ref{fig:NS_collapse}. We also show the evolution 
of the lapse (along the $x$-axes) for various time slices in Fig.~\ref{fig:Lapse_collapse}. This 
"collapse of the lapse" is characteristic for the formation of a  black hole.

\begin{figure}
   \centering
   \hspace*{-0.15cm}\includegraphics[width=1.\columnwidth]{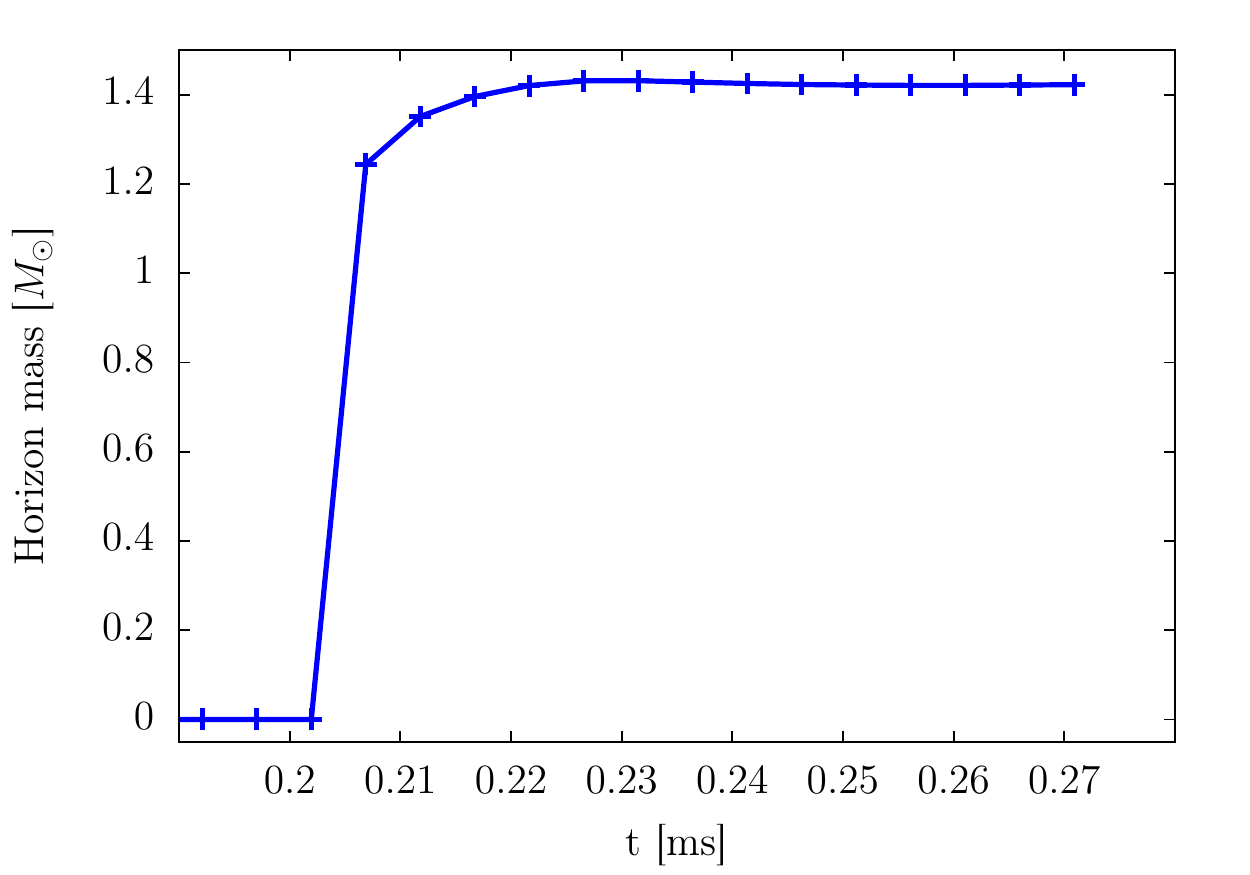}
   \caption{Evolution of the apparent horizon mass for the test with a collapsing neutron star.}
\label{fig:AH_mass}
\end{figure}

In order to make sure, that the region where we remove particles (defined by
the lapse threshold $\alpha_{\rm thr}$) is well contained within the
event horizon, we  imported our metric data into the Einstein
Toolkit \cite{ETK:web,Loffler:2011ay}  in order to apply the
apparent horizon finder \Ahf of  \cite{Thornburg:1995cp}
to our data, see Fig.~\ref{fig:AH_mass}. We find an apparent horizon
for the first time shortly after  0.205 ms (the zero values before that 
simply indicate that no apparent horizon was  found) with an irreducible mass 
close to  1.24 \Msun which grows initially, reaches a maximum value of
1.4314 \msun, decreases slightly and then starts to increase again until the
end of the simulation.

\begin{figure}
   \hspace*{-0.7cm}\includegraphics[width=1.2\columnwidth]{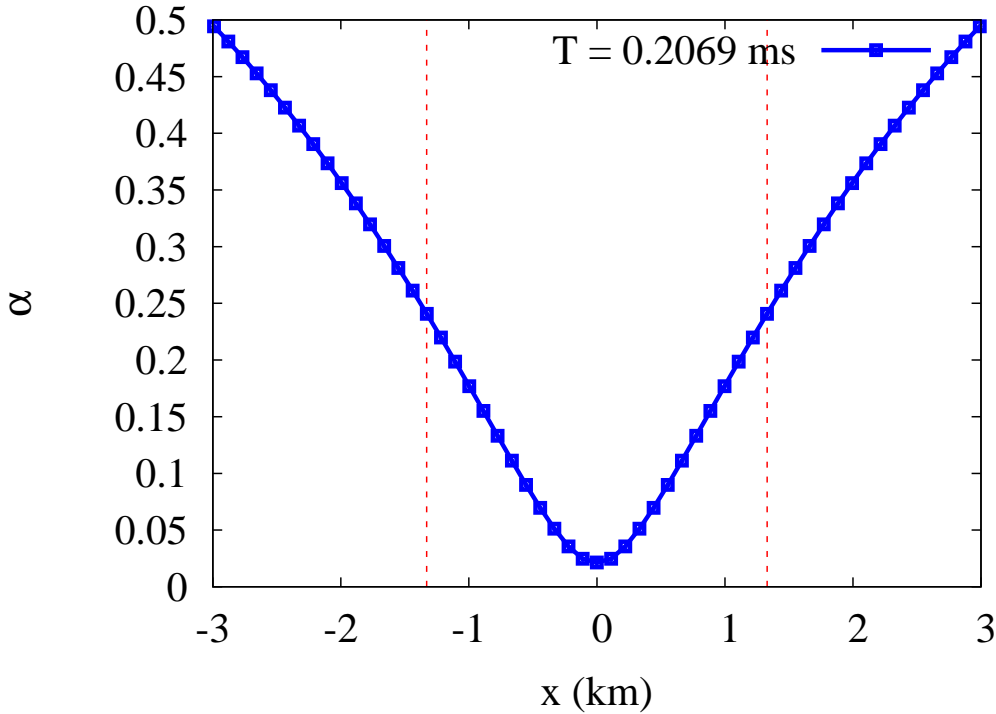}
   \caption{Lapse profile at the time of first finding the apparent horiozn.
            The vertical dashed lines show the size of the apparent horizon.
            The squares indicate the location of our grid points.}
\label{fig:AH_lapse}
\end{figure}

In Fig.~\ref{fig:AH_lapse} we show the lapse profile at the time when the
apparent horizon is first found. The vertical dashed red lines show the
location of the apparent horizon. As the event horizon is guaranteed to
be outside the apparent horizon at all times, we do have enough
grid resolution, so that the removal of SPH particles (where
$\alpha < \alpha_{\rm thr}$) can not  affect the region outside the
horizon. As can be seen from Fig.~16 in \cite{Baiotti:2004wn} the event
horizon forms typically up to 1 ms before the apparent horizon and grows
rapidly in size. Therefore, the fact that we start removing particles slightly
before ($\approx$ 0.01 ms) the apparent horizon forms, is not a cause for
concern. All particles were significantly inside of the horizon at the
time of their removal.

\section{Summary and conclusions}
\label{sec:summary}
In this paper we have presented the methodology behind what, to the best of our knowledge,
is the first Lagrangian fully General Relativistic hydrodynamics code. Part of the motivation 
for our alternative approach comes from the recent breakthrough in multi-messenger astrophysics
where a neutron star merger was observed both in gravitational and electromagnetic waves.
While the gravitational waves are dominated by the bulk matter motion in the densest, innermost
regions of the remnant, the electromagnetic emission is caused by comparatively small amounts
of matter that are ejected from the merger site. Such ejecta pose a serious challenge
to Eulerian methods, but are comparably straight forward to  evolve in a Lagrangian approach
such as ours.\\
In our new  code \SpB we evolve the matter by means of
Lagrangian particles according to a General Relativistic Smoothed Particle Hydrodynamics (SPH)
formulation, see Sec.~\ref{sec:hydro}. This formulation profits from a number of recent major improvements that have been
discussed and extensively tested in a non-relativistic context and implemented in the \Ma code
\cite{rosswog20a}. The improvements include the use of high order Wendland kernel 
functions, a slope-limited  reconstruction in the dissipation tensor, and a novel steering of the 
artificial dissipation by means of monitoring the exact conservation of entropy  \cite{rosswog20b}.\\
Relativistic gravity enters the fluid equations of motion via (derivatives  and the determinant of) 
the metric tensor. We evolve the metric, see Sec.~\ref{sec:spacetime_evolution},  like in most Eulerian hydrodynamics formulations
by following the Baumgarte-Shapiro-Shibata-Nakamura (BSSN) approach  \cite{nakamura87,shibata95,baumgarte99}. 
For now, we solve the BSSN equations on a uniform Cartesian grid and calculate derivates 
via finite differencing (of either 4th, 6th or 8th order). \\
An important element of our approach is the coupling between the fluid (on particles) and
the spacetime (known on a mesh), see Sec.~\ref{sec:particle_mesh}. At every (sub-)step the energy-momentum tensor
$T_{\mu\nu}$ of the fluid needs to be mapped from the particles to the grid points ($P2M$) 
and the  metric tensor properties need to be mapped back to the particle positions ($M2P$). 
Both of these steps turned out to be crucial for the accuracy of our scheme. We found in 
particular that a straight forward mapping with SPH kernels in the $P2M$-step was not a 
good choice. Instead, we used a number of more accurate (but not strictly positive definite)
kernels that have been developed in the context of vortex-particle methods. In addition, we have implemented
a mapping via a Moving Least Square (MLS) method which requires the frequent solution of 
$10\times10$ equation system. While this comes at some computational cost, it is
still acceptable in the overall computational balance.
Also for the $M2P$-step we have implemented several options including a
weighted essentially non-oscillatory interpolation of order 5 (WENO5) and a
5th-order Hermite interpolation (inspired by and extending the work of Timmes and Swesty \cite{timmes00a}).
We found several mapping combinations to work well and we have chosen as defaults
the MLS method in the $P2M$ and the 5th-order Hermite interpolation in the $M2P$ step.\\
A number of our test cases involves neutron stars which we set up 
according to a relativistic version of the "Artificial Pressure Method" (APM) that has recently been suggested in
a Newtonian context \cite{rosswog20a}. Its main purpose is to obtain a particle distribution
that reflects a given density profile, see Sec.~\ref{sec:APM}. Starting from some initial SPH particle 
distribution, an iteration is performed that drives the particles into
locations where they minimize their density error. This is achieved by  an Euler-type
equation which uses an "artificial pressure" that is based on a local density error measure.\\
We have scrutinized our methodology and implementation via a number of standard tests
that are often used for Eulerian Numerical Relativity codes, see Sec.~\ref{sec:tests}. All tests were performed with the
full 3+1 code. We have tested the {\em special}-relativistic hydrodynamics part via a relativistic
shock tube benchmark, the {\em general-}relativistic hydrodynamics  by evolving a neutron star
in a fixed spacetime ("Cowling approximation") and, in a next step, we evolved the
hydrodynamics together with the metric. 
In both of the latter cases the neutron stars remain very close
to the exact TOV-solution and they oscillate at frequencies that are in excellent agreement
with those found in semi-analytic and Eulerian approaches. Contrary to the latter approaches,
the neutron star surface does not pose any challenge for \spB, it remains sharp throughout the
evolution, see e.g. Fig.~\ref{fig:TOV_2.0_0_10ms}, and does not require any special treatment. 
We further present our results for the challenging "migration test" where, triggered by truncation 
error alone, an unstable neutron star transitions via violent oscillations into a stable configuration. 
And finally, when a small velocity perturbation is added to the same neutron star, it collapses and forms
a black hole. In all of these tests, we obtain results that are in very good agreement with those of
established Eulerian Numerical Relativity codes.

\section*{Acknowledgments}

It is a pleasure to acknowledge stimulating and insightful conversations with Luis Lehner
in the early phase of this project and continued discussions with Vivek Chaurasia and 
in particular Francesco Torsello in the mature phases of this project. We also want to thank
Ian Hawke for comments on the first archive version of the paper.  SR has been 
supported by the Swedish Research Council (VR) under grant number 2016- 03657\_3, by 
the Swedish National Space Board under grant number  Dnr. 107/16,  by the 
research environment grant ``Gravitational Radiation and Electromagnetic Astrophysical
Transients (GREAT)'' funded by the Swedish Research 
council (VR) under Dnr 2016-06012 and by the Knut and Alice Wallenberg Foundation
under grant Dnr. KAW 2019.0112. We gratefully 
acknowledge inspiring interactions via the COST Action CA16104 
``Gravitational waves, black holes and fundamental physics'' (GWverse) and  COST Action CA16214
``The multi-messenger physics and astrophysics of neutron stars'' (PHAROS).
PD would like to thank the Astronomy Department at SU and the Oscar Klein Centre for their hospitality during
numerous visits in the course of the development of \spB.
The simulations for this paper were performed on the facilities of the North-German Supercomputing Alliance (HLRN),
and on the resources provided by the Swedish National Infrastructure for Computing (SNIC) 
in Link\"oping  partially funded by the Swedish Research Council through grant agreement no. 2016-07213. 
Portions of this research were also conducted with high
performance computational resources provided by the Louisiana Optical Network
Infrastructure (http://www.loni.org).
 \\
 
\section{Appendix A}
 The so-called "W-method" \cite{marronetti08,tichy07} also starts from the ADM variables 
$\gamma_{ij}$, $K_{ij}$, $\alpha$ and
$\beta^{i}$ and  defines the BSSN variables as
\begin{eqnarray}
  W & = & \gamma^{-1/6}, \\
  \tlg_{ij} & = & \emfpW \gamma_{ij}, \\
  K & = & \gamma^{ij} K_{ij}, \\
  \tlG^{i} & = & \tlg^{jk} \tlG^{i}_{jk}, \\
  \tlA_{ij} & = & \emfpW\left ( K_{ij}-\frac{1}{3}\gamma_{ij} K\right ).
\end{eqnarray}
These quantities are evolved according to 
\begin{eqnarray}
  \dt{W} & = & \frac{1}{3} W \left ( \alpha K - \pdu{\beta}{i}{i} \right) + \upwindu{W}{}{i}, \\
  \dt{\gamma_{ij}} & = & -2\alpha \tlA_{ij} + \tlg_{ik} \pdu{\beta}{k}{j}
                         + \tlg_{jk} \pdu{\beta}{k}{i} \nonumber\\
                       & &  -\frac{2}{3} \tlg_{ij} \pdu{\beta}{k}{k}
                         + \upwindl{\tlg}{ij}{k}, \\
  \dt{K} & = & -\emfpW \left ( \tlg^{ij} \left [ \pdpdu{\alpha}{}{i}{j}
               -\frac{1}{W}\pdu{W}{}{i}\pdu{\alpha}{}{j} \right ] 
               - \tlGn^{i}\pdu{\alpha}{}{i} \right ) \nonumber \\
         &   & + \alpha \left ( \tlA^{i}_{j} \tlA^{j}_{i} +\frac{1}{3} K^2
               \right ) + \upwindu{K}{}{i} + 4 \pi \alpha ( \rho + s ), \\
  \dt{\tlG^{i}} & = & -2 \tlA^{ij} \pdu{\alpha}{}{j} +\tlg^{jk} \pdpdu{\beta}{i}{j}{k} + \frac{1}{3}
                    \tlg^{ij} \pdpdu{\beta}{k}{j}{k}  \nonumber\\
 & &+  2 \alpha \left (
                    \tlG^{i}_{jk} \tlA^{jk} - \frac{2}{3} \tlg^{ij}
                    \pdu{K}{}{j} - \frac{3}{W} \tlA^{ij} \pdu{W}{}{j}\right ) 
                    \nonumber \\
                &   &  -\tlGn^{j}\pdu{\beta}{i}{j}
                    + \frac{2}{3} \tlGn^{i}\pdu{\beta}{j}{j} 
                + \upwindu{\tlG}{i}{j} \nonumber\\
                & & -16 \pi \alpha \tlg^{ij} s_j, \\
  \dt{\tlA_{ij}} & = & \emfpW [ -\pdpdu{\alpha}{}{i}{j} + \tlG^{k}{ij}
                       \pdu{\alpha}{}{k} +\alpha R_{ij} \nonumber\\
                      & &- 
                     ( \pdu{\alpha}{}{i}
                       \pdu{W}{}{j}+\pdu{\alpha}{}{j} \pdu{W}{}{i} )
                       ]^{\mathrm{TF}} \nonumber \\
                 &   & +\alpha ( K \tlA_{ij}- 2 \tlA_{ik} \tlA^{k}_{j} )
                       + \tlA_{ik} \pdu{\beta}{k}{j} \nonumber \\
                  & &     + \tlA_{jk} \pdu{\beta}{k}{i}
                       - \frac{2}{3} \tlA_{ij} \pdu{\beta}{k}{k}  +\upwindl{\tlA}{ij}{k}  \nonumber \\
                 &   &  - \emfpW \alpha 8 \pi
                       \left (T_{ij}-\frac{1}{3} \gamma_{ij} s\right ),
\end{eqnarray}
where $\rho$, $s$ and $s_i$ are given by Eqs.~(\ref{eq:BSSN_rho})-(\ref{eq:BSSN_Si}).
Finally $R_{ij} = \tlR_{ij} + R^{W}_{ij}$, where $\tlR_{ij}$ is given as for the 
$\phi-$method, while
\begin{eqnarray}
 R^{W}_{ij} & = & \frac{1}{W}\pdpdu{W}{}{i}{j}-\frac{1}{W^2}\pdu{W}{}{i}\pdu{W}{}{j}
             +2\tlG^{k}_{ij}\pdu{W}{}{k} \nonumber \\
       &   & +\tlg_{ij}\tlg^{kl}\left (\frac{1}{W}\pdpdu{W}{}{i}{j}
             -\frac{1}{W^2}\pdu{W}{}{i}\pdu{W}{}{j}
             +2\tlG^{k}_{ij}\pdu{W}{}{k} \right ) \nonumber \\
       &   & +\frac{1}{W^2}\pdu{W}{}{i}\pdu{W}{}{j}
             -\tlg_{ij}\tlg^{kl}\frac{1}{W^2}\pdu{W}{}{k}\pdu{W}{}{l}.
\end{eqnarray}

\bibliographystyle{unsrt}
\bibliography{astro_SKR_SPHINCS_BSSN}

\label{lastpage}

\end{document}